\documentclass[12pt, oneside,a4paper]{article}

\usepackage{amsmath, amsthm, amssymb,amsfonts}
\usepackage{bbm, bm, booktabs}
\usepackage{braket}
\usepackage{calrsfs, color, caption}
\usepackage{esint, enumitem}
\usepackage{fancyhdr, float}
\usepackage{graphicx, graphics, geometry}
\usepackage{indentfirst}
\usepackage{algorithm}
\usepackage{algorithmicx} 
\usepackage{algpseudocode}
\usepackage{listings}
\usepackage{multirow}
\definecolor{dkgreen}{rgb}{0, 0.6, 0}
\definecolor{gray}{rgb}{0.5, 0.5, 0.5}
\definecolor{mauve}{rgb}{0.58, 0, 0.82}
\usepackage[T1]{fontenc}
\usepackage[labelfont=bf]{caption}
\usepackage{stmaryrd, siunitx, subfigure ,bbding}
\usepackage[stable]{footmisc}
\usepackage{tikz, textcomp}
\usetikzlibrary{fit, positioning, arrows, automata}
\usepackage{verbatim}
\usepackage{wasysym}
\usepackage{wrapfig}
\usepackage{framed}
\usepackage{extarrows}
\usepackage{hyperref}
\usepackage{adjustbox}
\usepackage{natbib}
\setcitestyle{authoryear, round}

\usepackage[cal=cm, scr=rsfs]{mathalfa}

\usepackage{setspace}
\DeclareMathOperator*{\argmin}{\arg\!\min}
 
\theoremstyle{plain}
\newtheorem{thm}{Theorem}

\newtheorem{prop}{Proposition} 
\newtheorem{cor}{Corollary} 
\theoremstyle{remark}
\newtheorem{assump}{Assumption} 
 
\newtheorem{rmk}{Remark} 
\newtheorem{condition}{Condition}

\date{}

\usepackage{authblk} 
\title{Collaborative Inference for Sparse High-Dimensional Models with Non-Shared Data}
\author[1]{Yifan Gu}      
\author[1]{Hanfang Yang}     
\author[1]{Songshan Yang}   
\author[2]{Hui Zou}   

\affil[1]{Renmin University of China} 
\affil[2]{University of Minnesota}  

\usepackage{titling}
\pretitle{\begin{center}\Large\bfseries}
\posttitle{\par\end{center}}
\preauthor{\begin{center}\normalsize}
\postauthor{\par\end{center}}
\predate{\begin{center}\small}
\postdate{\par\end{center}}
\begin{document}
%
%
\newcommand{\trans}[0]{{\mathrm{T}}}
\newcommand{\dd}[0]{{\mathrm{d}}}
\newcommand{\lambdada}[0]{{\boldsymbol{\lambda}}}
\newcommand{\betada}[0]{{\boldsymbol{\beta}}}
\newcommand{\deltada}[0]{{\boldsymbol{\delta}}}
\newcommand{\Jda}[0]{{\mathbf{J}}}
\newcommand{\Cda}[0]{{\mathbf{C}}}
\newcommand{\omegada}[0]{{\boldsymbol{\omega}}}
\newcommand{\vda}[0]{{\boldsymbol{v}}}
\newcommand{\zda}[0]{{\boldsymbol{z}}}
\newcommand{\uda}[0]{{\boldsymbol{u}}}
\newcommand{\xda}[0]{{\boldsymbol{x}}}
\newcommand{\alphada}[0]{{\boldsymbol{\alpha}}}
\newcommand{\iternum}[1]{{(#1)}}
\newcommand{\normtwo}[1]{ \Vert#1\Vert_{2} }
\newcommand{\norminf}[1]{ \Vert#1\Vert_{\infty}} 
\newcommand{\normmin}[1]{{\Vert#1\Vert_{\min}}}
\newcommand{\normone}[1]{\Vert#1\Vert_{1}}
\newcommand{\ora}[0]{{\mathrm{ora}}}
\newcommand{\JdaS}[0]{{\Jda_{\mathcal{S}}}}
\newcommand{\yda}[0]{{\boldsymbol{y}}}
\newcommand{\MC}[0]{{{\mathcal{M}^c}}}
\newcommand{\tda}[0]{{\boldsymbol{t}}}
\newcommand{\dda}[0]{{\boldsymbol{d}}}
 \newcommand{\normome}[1]{{\Vert#1\Vert_{\boldsymbol{\Omega}}}}
\newcommand{\JdaSM}[0]{{\Jda_{\os}}}
\newcommand{\Jdaos}[0]{{\Jda_{\overline{\mathcal{S}}}}}
\newcommand{\Cmat}[0]{{\begin{pmatrix}\Cda^\trans\\ \zeroda_{r\times s}^\trans\end{pmatrix}}}
\newcommand{\Cmatsingle}[0]{{\begin{pmatrix}1\\ \zeroda_{1\times s}^\trans\end{pmatrix}}}
\newcommand{\zeroda}[0]{{\mathbf{0}}}
\newcommand{\hda}[0]{{\boldsymbol{h}}}
\newcommand{\betadagger}[0]{{\betada^\dagger}}
\newcommand{\thetadagger}[0]{{\btheta^\dagger}}
\newcommand{\nablatheta}[0]{{\nabla_{\btheta}}}
\newcommand{\nablagamma}[0]{{\nabla_{\bgamma}}}
\newcommand{\hmL}[0]{{\widehat{\mathcal{L}}}}
\newcommand{\mathL}[0]{\mathcal{L}}
\newcommand{\otc}[0]{{  \mathcal{T}_a^c}}
\newcommand{\ot}[0]{{ \mathcal{T}_a}}
\newcommand{\osc}[0]{{ \mathcal{S}_a^c}}
\newcommand{\os}[0]{{ \mathcal{S}_a}} 
\newcommand{\Tgamma}[0]{{{\mathcal{T}}}}
\newcommand{\Sgamma}[0]{{{\mathcal{S}}}}
\newcommand{\Tgammac}[0]{{{\mathcal{T}^c}}}
\newcommand{\Sgammac}[0]{{{\mathcal{S}^c}}}
\newcommand{\normmax}[1]{{\Vert#1\Vert_{\max}}}
\newcommand{\nablabb}[0]{\nabla_{\boldsymbol{b}}}
\newcommand{\bda}[0]{{\boldsymbol{b}}}
\newcommand{\obC}[0]{  \bC_a}  
%
%
\def\bzero{{\bf 0}}
\def\bone{{\bf 1}}
%
%
%
%
\def\ba{{\mbox{\boldmath$a$}}}
\def\bb{{\bf b}}
\def\bc{{\bf c}}
\def\bd{{\bf d}}
\def\be{{\bf e}}
\def\bdf{{\bf f}}
\def\bg{{\mbox{\boldmath$g$}}}
\def\bh{{\bf h}}
\def\bi{{\bf i}}
\def\bj{{\bf j}}
\def\bk{{\bf k}}
\def\bl{{\bf l}}
\def\bm{{\bf m}}
\def\bn{{\bf n}}
\def\bo{{\bf o}}
\def\bp{{\bf p}}
\def\bq{{\bf q}}
\def\br{{\bf r}}
\def\bs{{\bf s}}
\def\bt{{\bf t}}
\def\bu{{\bf u}}
\def\bv{{\bf v}}
\def\bw{{\bf w}}
\def\bx{{\bf x}}
\def\by{{\bf y}}
\def\bz{{\bf z}}
\def\bA{{\bf A}}
\def\bB{{\bf B}}
\def\bC{{\bf C}}
\def\bD{{\bf D}}
\def\bE{{\bf E}}
\def\bF{{\bf F}}
\def\bG{{\bf G}}
\def\bH{{\bf H}}
\def\bI{{\bf I}}
\def\bJ{{\bf J}}
\def\bK{{\bf K}}
\def\bL{{\bf L}}
\def\bM{{\bf M}}
\def\bN{{\bf N}}
\def\bO{{\bf O}}
\def\bP{{\bf P}}
\def\bQ{{\bf Q}}
\def\bR{{\bf R}}
\def\bS{{\bf S}}
\def\bT{{\bf T}}
\def\bU{{\bf U}}
\def\bV{{\bf V}}
\def\bW{{\bf W}}
\def\bX{{\bf X}}
\def\bY{{\bf Y}}
\def\bZ{{\bf Z}}
\def\smbZ{\scriptstyle{\bf Z}}
\def\smM{\scriptstyle{M}}
\def\smN{\scriptstyle{N}}
\def\smbT{\scriptstyle{\bf T}}
%
%
%
%
\def\thick#1{\hbox{\rlap{$#1$}\kern0.25pt\rlap{$#1$}\kern0.25pt$#1$}}
\def\balpha{\boldsymbol{\alpha}}
\def\bbeta{\boldsymbol{\beta}}
\def\bgamma{\boldsymbol{\gamma}}
\def\bdelta{\boldsymbol{\delta}}
\def\bepsilon{\boldsymbol{\epsilon}}
\def\bvarepsilon{\boldsymbol{\varepsilon}}
\def\bzeta{\boldsymbol{\zeta}}
\def\bdeta{\boldsymbol{\eta}}
\def\btheta{\boldsymbol{\theta}}
\def\biota{\boldsymbol{\iota}}
\def\bkappa{\boldsymbol{\kappa}}
\def\blambda{\boldsymbol{\lambda}}
\def\bmu{\boldsymbol{\mu}}
\def\bnu{\boldsymbol{\nu}}
\def\bxi{\boldsymbol{\xi}}
\def\bomicron{\boldsymbol{\omicron}}
\def\bpi{\boldsymbol{\pi}}
\def\brho{\boldsymbol{\rho}}
\def\bsigma{\boldsymbol{\sigma}}
\def\btau{\boldsymbol{\tau}}
\def\bupsilon{\boldsymbol{\upsilon}}
\def\bphi{\boldsymbol{\phi}}
\def\bchi{\boldsymbol{\chi}}
\def\bpsi{\boldsymbol{\psi}}
\def\bomega{\boldsymbol{\omega}}
\def\bAlpha{\boldsymbol{\Alpha}}
\def\bBeta{\boldsymbol{\Beta}}
\def\bGamma{\boldsymbol{\Gamma}}
\def\bDelta{\boldsymbol{\Delta}}
\def\bEpsilon{\boldsymbol{\Epsilon}}
\def\bZeta{\boldsymbol{\Zeta}}
\def\bEta{\boldsymbol{\Eta}}
\def\bTheta{\boldsymbol{\Theta}}
\def\bIota{\boldsymbol{\Iota}}
\def\bKappa{\boldsymbol{\Kappa}}
\def\bLambda{{\boldsymbol{\Lambda}}}
\def\bMu{\boldsymbol{\Mu}}
\def\bNu{\boldsymbol{\Nu}}
\def\bXi{\boldsymbol{\Xi}}
\def\bOmicron{\boldsymbol{\Omicron}}
\def\bPi{\boldsymbol{\Pi}}
\def\bRho{\boldsymbol{\Rho}}
\def\bSigma{\boldsymbol{\Sigma}}
\def\bTau{\boldsymbol{\Tau}}
\def\bUpsilon{\boldsymbol{\Upsilon}}
\def\bPhi{\boldsymbol{\Phi}}
\def\bChi{\boldsymbol{\Chi}}
\def\bPsi{\boldsymbol{\Psi}}
\def\bOmega{\boldsymbol{\Omega}}

\def\smalpha{{{\scriptstyle{\alpha}}}}
\def\smbeta{{{\scriptstyle{\beta}}}}
\def\smgamma{{{\scriptstyle{\gamma}}}}
\def\smdelta{{{\scriptstyle{\delta}}}}
\def\smepsilon{{{\scriptstyle{\epsilon}}}}
\def\smvarepsilon{{{\scriptstyle{\varepsilon}}}}
\def\smzeta{{{\scriptstyle{\zeta}}}}
\def\smdeta{{{\scriptstyle{\eta}}}}
\def\smtheta{{{\scriptstyle{\theta}}}}
\def\smiota{{{\scriptstyle{\iota}}}}
\def\smkappa{{{\scriptstyle{\kappa}}}}
\def\smlambda{{{\scriptstyle{\lambda}}}}
\def\smmu{{{\scriptstyle{\mu}}}}
\def\smnu{{{\scriptstyle{\nu}}}}
\def\smxi{{{\scriptstyle{\xi}}}}
\def\smomicron{{{\scriptstyle{\omicron}}}}
\def\smpi{{{\scriptstyle{\pi}}}}
\def\smrho{{{\scriptstyle{\rho}}}}
\def\smsigma{{{\scriptstyle{\sigma}}}}
\def\smtau{{{\scriptstyle{\tau}}}}
\def\smupsilon{{{\scriptstyle{\upsilon}}}}
\def\smphi{{{\scriptstyle{\phi}}}}
\def\smchi{{{\scriptstyle{\chi}}}}
\def\smpsi{{{\scriptstyle{\psi}}}}
\def\smomega{{{\scriptstyle{\omega}}}}
\def\smAlpha{{{\scriptstyle{\Alpha}}}}
\def\smBeta{{{\scriptstyle{\Beta}}}}
\def\smGamma{{{\scriptstyle{\Gamma}}}}
\def\smDelta{{{\scriptstyle{\Delta}}}}
\def\smEpsilon{{{\scriptstyle{\Epsilon}}}}
\def\smZeta{{{\scriptstyle{\Zeta}}}}
\def\smEta{{{\scriptstyle{\Eta}}}}
\def\smTheta{{{\scriptstyle{\Theta}}}}
\def\smIota{{{\scriptstyle{\Iota}}}}
\def\smKappa{{{\scriptstyle{\Kappa}}}}
\def\smLambda{{{\scriptstyle{\Lambda}}}}
\def\smMu{{{\scriptstyle{\Mu}}}}
\def\smNu{{{\scriptstyle{\Nu}}}}
\def\smXi{{{\scriptstyle{\Xi}}}}
\def\smOmicron{{{\scriptstyle{\Omicron}}}}
\def\smPi{{{\scriptstyle{\Pi}}}}
\def\smRho{{{\scriptstyle{\Rho}}}}
\def\smSigma{{{\scriptstyle{\Sigma}}}}
\def\smTau{{{\scriptstyle{\Tau}}}}
\def\smUpsilon{{{\scriptstyle{\Upsilon}}}}
\def\smPhi{{{\scriptstyle{\Phi}}}}
\def\smChi{{{\scriptstyle{\Chi}}}}
\def\smPsi{{{\scriptstyle{\Psi}}}}
\def\smOmega{{{\scriptstyle{\Omega}}}}

\def\smbalpha{\boldsymbol{{\scriptstyle{\alpha}}}}
\def\smbbeta{\boldsymbol{{\scriptstyle{\beta}}}}
\def\smbgamma{\boldsymbol{{\scriptstyle{\gamma}}}}
\def\smbdelta{\boldsymbol{{\scriptstyle{\delta}}}}
\def\smbepsilon{\boldsymbol{{\scriptstyle{\epsilon}}}}
\def\smbvarepsilon{\boldsymbol{{\scriptstyle{\varepsilon}}}}
\def\smbzeta{\boldsymbol{{\scriptstyle{\zeta}}}}
\def\smbdeta{\boldsymbol{{\scriptstyle{\eta}}}}
\def\smbtheta{\boldsymbol{{\scriptstyle{\theta}}}}
\def\smbiota{\boldsymbol{{\scriptstyle{\iota}}}}
\def\smbkappa{\boldsymbol{{\scriptstyle{\kappa}}}}
\def\smblambda{\boldsymbol{{\scriptstyle{\lambda}}}}
\def\smbmu{\boldsymbol{{\scriptstyle{\mu}}}}
\def\smbnu{\boldsymbol{{\scriptstyle{\nu}}}}
\def\smbxi{\boldsymbol{{\scriptstyle{\xi}}}}
\def\smbomicron{\boldsymbol{{\scriptstyle{\omicron}}}}
\def\smbpi{\boldsymbol{{\scriptstyle{\pi}}}}
\def\smbrho{\boldsymbol{{\scriptstyle{\rho}}}}
\def\smbsigma{\boldsymbol{{\scriptstyle{\sigma}}}}
\def\smbtau{\boldsymbol{{\scriptstyle{\tau}}}}
\def\smbupsilon{\boldsymbol{{\scriptstyle{\upsilon}}}}
\def\smbphi{\boldsymbol{{\scriptstyle{\phi}}}}
\def\smbchi{\boldsymbol{{\scriptstyle{\chi}}}}
\def\smbpsi{\boldsymbol{{\scriptstyle{\psi}}}}
\def\smbomega{\boldsymbol{{\scriptstyle{\omega}}}}
\def\smbAlpha{\boldsymbol{{\scriptstyle{\Alpha}}}}
\def\smbBeta{\boldsymbol{{\scriptstyle{\Beta}}}}
\def\smbGamma{\boldsymbol{{\scriptstyle{\Gamma}}}}
\def\smbDelta{\boldsymbol{{\scriptstyle{\Delta}}}}
\def\smbEpsilon{\boldsymbol{{\scriptstyle{\Epsilon}}}}
\def\smbZeta{\boldsymbol{{\scriptstyle{\Zeta}}}}
\def\smbEta{\boldsymbol{{\scriptstyle{\Eta}}}}
\def\smbTheta{\boldsymbol{{\scriptstyle{\Theta}}}}
\def\smbIota{\boldsymbol{{\scriptstyle{\Iota}}}}
\def\smbKappa{\boldsymbol{{\scriptstyle{\Kappa}}}}
\def\smbLambda{\boldsymbol{{\scriptstyle{\Lambda}}}}
\def\smbMu{\boldsymbol{{\scriptstyle{\Mu}}}}
\def\smbNu{\boldsymbol{{\scriptstyle{\Nu}}}}
\def\smbXi{\boldsymbol{{\scriptstyle{\Xi}}}}
\def\smbOmicron{\boldsymbol{{\scriptstyle{\Omicron}}}}
\def\smbPi{\boldsymbol{{\scriptstyle{\Pi}}}}
\def\smbRho{\boldsymbol{{\scriptstyle{\Rho}}}}
\def\smbSigma{\boldsymbol{{\scriptstyle{\Sigma}}}}
\def\smbTau{\boldsymbol{{\scriptstyle{\Tau}}}}
\def\smbUpsilon{\boldsymbol{{\scriptstyle{\Upsilon}}}}
\def\smbPhi{\boldsymbol{{\scriptstyle{\Phi}}}}
\def\smbChi{\boldsymbol{{\scriptstyle{\Chi}}}}
\def\smbPsi{\boldsymbol{{\scriptstyle{\Psi}}}}
\def\smbOmega{\boldsymbol{{\scriptstyle{\Omega}}}}
\newcommand{\mD}{\mathcal{D}} 
\newcommand{\mE}{\mathcal{E}}
\newcommand{\mG}{\mathcal{G}} 
\newcommand{\mP}{\mathcal{P}}
\newcommand{\mX}{\mathcal{X}}
\newcommand{\mY}{\mathcal{Y}}
\newcommand{\mU}{\mathcal{U}}  
\newcommand{\mZ}{\mathcal{Z}} 
\newcommand{\mI}{\mathcal{I}} 
\newcommand{\mR}{\mathcal{R}}
\newcommand{\mS}{\mathcal{S}}
\newcommand{\mM}{\mathcal{M}}
\newcommand{\mQ}{\mathcal{Q}}


\maketitle
\vspace{-1cm}

\noindent{\textbf{Abstract:}}
In modern data analysis, statistical efficiency improvement is expected via effective collaboration among multiple data holders with non-shared data. In this article, we propose a collaborative score-type test (CST) for testing linear hypotheses, which accommodates potentially high-dimensional nuisance parameters and a diverging number of constraints and target parameters. Through a careful decomposition of the Kiefer-Bahadur representation for the traditional score statistic, we identify and approximate the key components using aggregated local gradient information from each data source. In addition, we employ a two-stage partial penalization strategy to shrink the approximation error and mitigate the bias from the high-dimensional nuisance parameters. Unlike existing methods, the CST procedure involves constrained optimization under non-shared and high-dimensional data settings, which requires novel theoretical developments. We derive the limiting distributions for the CST statistic under the null hypothesis and the local alternatives. Besides, the CST exhibits an oracle property and achieves the global statistical efficiency.  Moreover, it relaxes the stringent restrictions on the number of data sources required in the current literature. Extensive numerical studies and a real example demonstrate the effectiveness and validity of our proposed method.  

\par \vspace{9pt} \noindent {\textbf{Keywords:} Collaborative statistical inference, Communication efficiency, High-dimensional testing, Linear hypothesis, Score test} 

\newpage

\section{Introduction}\label{sec:intro}

The modern paradigm of data collection poses new challenges for statistical inference. In many real-world applications, multiple data holders independently collect data for a common task. One prominent example is polygenic risk prediction in genome-wide association studies \citep{chatterjee2013projecting}, where many laboratories gather samples with a huge amount of genetic variants to predict the risk of complex traits or diseases. Given the high cost of genetic data collection, aggregating data from different labs can greatly enhance predictive power at a relatively low cost. However, direct data sharing gives rise to privacy and ownership-related concerns, while limited bandwidth and practical costs hinder communications between different sites. Therefore, it is essential to develop collaborative methods for multiple data holders with non-shared data, especially for statistical inference in high-dimensional models.

When facing non-shared data from multiple holders, one may think about a related problem called distributed computation of statistical methods, which has been studied by many authors. A list of references covering distributed estimation methods includes, but not limited to, \citet{lee2017communication}, \citet{battey2018distributed}, \citet{jordan2018communication}, \citet{wenxinzhou2022distributedhuber} and \citet{fan2021communication}. In this work our primary focus is hypothesis testing. Effective collaboration among multiple data holders is expected to improve the power of statistical discoveries. Nevertheless, research on distributed inference, especially in high-dimensional scenarios, remains limited due to the complexities of asymptotic expansions for estimators obtained through distributed procedures \citep{chen2021distributed}. For instance, \citet{jordan2018communication} developed a statistical inference method only for a fixed-dimensional model using a gradient-enhanced surrogate loss, which is difficult to extend to the high-dimensional scenario due to non-negligible bias. As an exception, \citet{battey2018distributed} explored a divide-and-conquer (DC) testing approach for sparse high-dimensional models based on an averaged debiased estimator. However, their method assumed balanced local sample sizes and imposed a stringent scaling on the number of data sources $m$. These limitations restrict broader collaboration across various data holders. For example, in modern Internet-of-Things networks, wearable devices automatically collect users' biological information for health event prediction \citep{chen2020fedhealth}. Treating the number of devices as negligible compared to the sample size can undermine the theoretical foundation of distributed inference methods. Moreover, divide-and-conquer begins with the assumption that one has access to all data and chooses to separate the whole data into multiple smaller batches. This is fundamentally different from the setting that we are considering in this work in which the original data are stored in multiple non-shared sites. Therefore, DC methods not only have theoretical limitations as discussed above but also have troubles to be applied in real applications with non-shared data.

To fill in the gap between practical concerns and statistical methods, we develop a communication-efficient collaborative inference procedure for a generic high-dimensional model when data are stored in multiple centers who are not allowed to share their data. Let the model parameter $\betada=(\btheta^\trans,\bgamma^\trans)^\trans$, where $\btheta\in\mathbb{R}^d$ and $\bgamma\in\mathbb{R}^{p-d}$ are the target parameter and potentially high-dimensional nuisance parameter, respectively. We focus on testing the linear hypothesis $H_0:\bC\btheta=\tda$, where $\bC\in\mathbb{R}^{r\times d}$ is the constraint matrix. The class of hypotheses we considered is more versatile than that considered by previous distributed inference literature in two aspects. First, we can test more complex hypotheses by designing the constraint matrix $\bC$, whereas previous studies like \citet{battey2018distributed} focused only on the univariate hypothesis. Second, we allow that both the number of constraints, $r$, and the dimensionality of the target parameter, $d$, can diverge at a certain rate with the sample size.

The proposed collaborative score-type test (CST) draws inspiration from the traditional score statistic used in low-dimensional scenarios with shared data. Through a careful decomposition of the Kiefer-Bahadur representation for the traditional score statistic, we identify and approximate its key components using aggregated local gradient information from each data source. This leads to the construction of a constrained surrogate collaborative loss, which incorporates a gradient shift without requiring direct data sharing. Interestingly, if we remove the linear constraint from our collaborative loss, then it coincides with the gradient-enhanced loss introduced by \citet{wang2017efficient}, \citet{jordan2018communication}, and \citet{fan2021communication}, though our derivation offers a novel perspective on its formulation. 

To ensure the validity of CST, we employ a two-stage partial penalization strategy to shrink the approximation error for the surrogate loss and mitigate the bias from the high-dimensional nuisance parameters.  The first stage brings the estimator into the neighborhood of the true parameter, while the second stage encodes the null hypothesis into the collaborative estimators with refined penalization on the high-dimensional nuisance parameters.  From a theoretical perspective, we introduce an oracle collaborative score-type test (OCST),  which takes the true support of high-dimensional nuisance parameter as prior knowledge. We demonstrate that the CST statistic possesses an oracle property, meaning that it shares the same limiting distribution as the OCST statistic. This allows us to use the tractable distribution of the OCST statistic to construct an asymptotically valid rejection region for CST. Recently, \citet{Liu18032025} studied the oracle property for a distributed sparse penalization method focused on unconstrained estimation. In contrast, we establish the oracle property of the collaborative inference procedure with linear constraints and high-dimensional nuisance parameters, which demands novel algorithmic and theoretical developments.

The CST achieves both statistical and communication efficiency. The power function of the CST aligns with that of the traditional score test in shared-data scenarios \citep{boos1992score}, showing the efficiency gains from collaboration over independent local tests.  Moreover, we demonstrate that the CST is more powerful than the DC test proposed by \cite{battey2018distributed}, since it circumvents the need for debiasing. Meanwhile, since the CST method only aggregates gradient information, it requires a communication budget of $O(mp)$ bits per iteration within the cluster. This is considered communication-efficient so long as the number of communication rounds remains reasonable \citep{fan2021communication}. We show that the increase in iterations generally scales logarithmically with the number of data sites $m$ and doubly logarithmically with the dimensionality $p$ for many classical regression problems. This scaling factor greatly reduces the communication overhead, enabling more data holders to cooperate in a distributed system. 

Overall, this article makes four major contributions. First, we propose the CST procedure for collaborative inference in a generic high-dimensional model with non-shared data. Second, we address the high-dimensional nuisance parameters by employing a two-stage partial penalization strategy. Third, compared with the prior work, we eliminate the stringent restrictions on the balanced sample sizes and the number of data sources. Besides, the linear hypothesis we considered is more general and flexible, allowing for a diverging number of constraints and target parameters. Fourth, we develop novel theoretical results for the constrained optimization with non-shared data, which is the core of the CST procedure that differs from existing approaches. In particular, a new oracle property for the CST statistic is established, providing a strong theoretical justification for our method.

The rest of this article is organized as follows. Section~\ref{sec:method} 
explains the heuristic motivation for the CST, and then introduces a two-stage partial penalization strategy to formally construct the CST statistic. Section~\ref{sec:theory} validates the CST procedure theoretically and studies its asymptotic power under general assumptions. The generalized linear model (GLM) is also considered as an illustrative example. Sections~\ref{sec:sim} provides a comprehensive numerical study, including simulation experiments and a real data example. We discuss some possible extensions in Section~\ref{sec:discuss}. 
All the proofs are deferred to the Supplementary Material.

\textbf{Notation}. We use bold uppercase letters to represent matrices. For $k \ge 2$, $\mathbf{I}_k$ and $\mathbf{O}_k$ represent an $k \times k$ identity matrix and zero matrix respectively. For any positive semi-definite matrix $\mathbf{A}$, we use $\lambda_{\max}(\mathbf{A})$ and $\lambda_{\min}(\mathbf{A})$ to denote its maximum and minimum eigenvalues.  For every integer $k \ge 1$, write $[k]=\{1, \ldots, k\}$. For a subset $\mathcal{J}$ of $[k]$ with cardinality $|\mathcal{J}|$, we write $\boldsymbol{u}_\mathcal{J} \in \mathbb{R}^{|\mathcal{J}|}$ as the subvector of $\boldsymbol{u}\in\mathbb{R}^k$ that consists of the entries indexed by $\mathcal{J}$. We use $\mathrm{supp}(\uda)=\{j\in[k]:u_j\neq0\}$ to denote the support of $\uda$. We use $\|\cdot\|_q$ $(1 \le q \le \infty)$ to denote the $\ell_q$-norm: $\|\boldsymbol{u}\|_q=(\sum_{i=1}^k\left|u_i\right|^q)^{1 / q}$ and $\|\boldsymbol{u}\|_{\infty}=\max _{1 \le i \le k}\left|u_i\right|$. $\normmin{\uda}$ denotes the minimum signal strength $\min_{j\in[k]}|u_j|$. We also use $\normmax{\mathbf{A}}$ and $\normtwo{\mathbf{A}}$ to denote the element-wise maximum norm and the operator norm induced by the $\ell_2$ norm  for any matrix $\mathbf{A}$. For any function $f: \mathbb{R} \mapsto \mathbb{R}$, we write $f(\boldsymbol{u})=\left(f\left(u_1\right), \ldots, f\left(u_k\right)\right)^{\mathrm{T}} \in \mathbb{R}^k$. Let $\Lambda(\mathtt{g})=\{\vda\in\mathbb{R}^p:\normone{\vda}\le \mathtt{g} \normtwo{\vda}\}$ and $\Theta(r)=\{\vda\in\mathbb{R}^p:\normtwo{\vda}\le r\}$ be the $\ell_1$ cone and $\ell_2$ ball in $\mathbb{R}^p$, respectively. Let $\chi^2(r,e)$ be a chi-square random variable with $r$ degrees of freedom and a non-centrality parameter $e$. For any two real numbers $u$ and $v$, we write $u \vee v=\max (u, v)$ and $u \wedge v=\min (u, v)$. For two sequences of non-negative numbers $\left\{a_n\right\}_{n \ge 1}$ and $\left\{b_n\right\}_{n \ge 1}$, $a_n \lesssim b_n$ indicates that $a_n \le C b_n$ for some constant $C>0$; $a_n \gtrsim b_n$ is equivalent to $b_n \lesssim a_n $; $a_n \asymp b_n$ is equivalent to $a_n \lesssim b_n$ and $b_n \lesssim a_n$.

\section{Collaborative score test}\label{sec:method}

\subsection{Preliminary}

Consider $m$ distinct sites, each with its own non-shared data and a sample size of $n_k$, leading to a collective sample size of $N=\sum_{k=1}^m n_k$. To be specific, for each $k=1, \ldots, m$, the $k$th local site stores a subset of $n_k$ observations, represented as $\mathcal{D}_k=\{\zda_i\}_{i \in \mathcal{I}_k}$, which is a random sample drawn from $\zda$ over some sample space $\mathcal{Z}$. Here, $\{\mathcal{I}_k\}_{k=1}^m$ corresponds to disjoint index sets that satisfy two conditions: (1) the union of all index sets covers the complete range of indices, i.e., $\cup_{k=1}^m \mathcal{I}_k=\{1, \ldots, N\}$, and (2) the cardinality of each set equals $n_k$, i.e. $|\mathcal{I}_k|=n_k$. 

For any parameter $\betada\in\mathbb{R}^p$, define its population risk function $\mathcal{L}(\betada)=\mathbb{E}_\zda \ell(\betada;\zda)$ based on a loss function $\ell(\cdot;\cdot):\mathbb{R}^p\times\mathcal{Z}$. We assume that $\ell(\betada;\zda)$ is convex and at least twice continuously differentiable in $\betada$. Under mild conditions, the population risk $\mathcal{L}$ is well-defined and has a unique minimizer $\betada^*$. In many statistical problems, the unknown parameter $\betada$ can be partitioned as $\betada=(\btheta^\trans,\bgamma^\trans)^\trans$, where $\btheta\in\mathbb{R}^d$ and $\bgamma\in\mathbb{R}^{p-d}$. Our goal is to conduct statistical inference on the parameter $\btheta^*$ through effective collaboration among different data sites while taking into consideration both the high-dimensional sparse nuisance parameter $\bgamma^*$ and the communication cost. In particular, we are interested in the following linear hypothesis
\begin{equation}\label{eq:linearH0}
 H_0:\, \mathbf{C}\btheta^* = \boldsymbol{t} \quad\text{vs.} \quad H_1:\, \mathbf{C}\btheta^* \neq \boldsymbol{t},
\end{equation}
for a given $r\times d$ constraint matrix $\mathbf{C}$ and an $r$-dimensional vector $\boldsymbol{t}$. We assume that $\mathbf{C}$ is of full row rank, implying that there are no redundant or contradictory constraints in $H_0$.

\subsection{Decomposition of the traditional score statistic}\label{subsec:decomposition}
To facilitate the discussions, we first introduce some notations.  Let $\Jda = \nabla^2\mathL(\betada^*)$ be the Hessian matrix of the risk function at the truth $\betada^*$,  $\obC=(\bC,\mathbf{O}_{r\times (p-d)})^\trans\in\mathbb{R}^{p\times r}$ be the augmented constraint matrix, and $\hda=\obC^\trans\betada^*-\tda=\bC\btheta^*-\tda$ be the local alternative parameter. Further define $\Psi = \obC^\trans\Jda^{-1}\obC$ and the projection matrix $\mathbf{P} =\Jda^{-1/2}\obC\Psi^{-1}\obC^\trans\Jda^{-1/2}$.

Although our focus is high-dimensional and non-shared data, we begin by examining the traditional score statistics in the low-dimensional setting with shared data. Given all data are accessible,  we construct the traditional score statistic by solving
$\widehat{\betada}^{\text{global}} = \argmin_{\bC\btheta=\tda} \widehat{\mathcal{L}}(\betada)$, where the global and local empirical loss functions are defined as
\[
\widehat{\mathcal{L}}(\betada) = \frac{1}{N}\sum_{k=1}^{m}n_k\widehat{\mathcal{L}}_{k}(\betada)=\frac{1}{N}\sum_{i=1}^{N}\ell(\betada;\zda_i) \quad \textrm{and} \quad \widehat{\mathcal{L}}_{k}(\betada) = \frac{1}{n_k}\sum_{i\in\mathcal{I}_k}\ell(\betada;\zda_i), \,\,\, k\in[m]. 
\]
The traditional score statistic is constructed based on $\nabla\widehat{\mathcal{L}}(\widehat{\betada}^{\text{global}})$, which is the empirical score function evaluated at the constrained estimator. To gain more intuition for the construction of the CST later, we investigate the asymptotic behavior of $\nabla\widehat{\mathcal{L}}(\widehat{\betada}^{\text{global}})$ through the decomposition
\[
\nabla\widehat{\mathcal{L}}(\widehat{\betada}^{\text{global}}) \approx \underbrace{\nabla\widehat{\mathcal{L}}({\betada}^{*}) +  \Jda (\widehat{\betada}^{\text{global}}-\betada^*)}_{I_1}+ \underbrace{(\nabla^2\widehat{\mathcal{L}}(\widetilde{\betada})-\Jda)(\widehat{\betada}^{\text{global}}-\betada^*)}_{I_2},
\]
where we use a heuristic ``Taylor's expansion'' of $\widehat{\mathcal{L}}(\cdot)$ and $\widetilde{\betada}$ lies in the line segment jointing $\widehat{\betada}^{\text{global}}$ and $\betada^*$. Under mild conditions, $I_2$ is of order $o(N^{-1/2})$. Thus we have
\begin{equation}\label{eq:tay}
\nabla\widehat{\mathcal{L}}(\widehat{\betada}^{\text{global}}) \approx I_1=\nabla\widehat{\mathcal{L}}({\betada}^{*}) +  \Jda (\widehat{\betada}^{\text{global}}-\betada^*).
\end{equation}
According to the Karush-Kuhn-Tucker (KKT) condition, there exists some $\widehat{\alphada}^{\text{global}}\in\mathbb{R}^r$ such that $\nabla\widehat{\mathcal{L}}(\widehat{\betada}^{\text{global}}) + \obC\widehat{\alphada}^{\text{global}} = \zeroda_p$. Hence, {using $\obC^\trans(\widehat{\betada}^{\text{global}}-\betada^*)=\tda-\bC\btheta^*=-\hda$ to solve for $\widehat{\alphada}^{\text{global}}$ from the equation $\obC\widehat{\alphada}^{\text{global}} =-\nabla\widehat{\mathcal{L}}({\betada}^{*})-\Jda (\widehat{\betada}^{\text{global}}-\betada^*)$}, we obtain the Kiefer-Bahadur representation for $\nabla\widehat{\mathcal{L}}(\widehat{\betada}^{\text{global}})$:
\[
\nabla\widehat{\mathcal{L}}(\widehat{\betada}^{\text{global}}) \approx \obC\Psi^{-1} \left(\obC^\trans\Jda^{-1}\nabla\hmL(\betada^*)-\hda\right).
\]
Since $\mathbb{E}[\nabla\ell(\betada^*;\zda_i)]=0$, we have that $\nabla\hmL(\betada^*)$ is the sum of $N$  independent and identically distributed random vectors with mean zero. Hence the asymptotic distribution of $\nabla\widehat{\mathcal{L}}(\widehat{\betada}^{\text{global}})$ becomes tractable. However, the global empirical loss $\widehat{\mathcal{L}}(\betada)$ is inaccessible in distributed scenarios as it requires completely shared data, which renders the global estimator $\widehat{\betada}^{\text{global}}$ invalid.  Note that the above analysis  extends to any surrogate loss function $\widetilde{\mathcal{L}}({\betada})$ and its constrained minimizer  $\widehat{\betada}=\argmin_{\bC\btheta=\tda} \widetilde{\mathcal{L}}({\betada})$. Following similar reasoning as in \eqref{eq:tay}, when $\widehat{\betada}$ is sufficiently close to $\betada^*$, we have
\[
\nabla\widehat{\mathcal{L}}(\widehat{\betada}) \approx 
 \nabla\widehat{\mathcal{L}}({\betada}^{*}) +  \Jda (\widehat{\betada}-\betada^*).
\]
It is then necessary to control the Kiefer-Bahadur representation for $\nabla\widehat{\mathcal{L}}(\widehat{\betada})$. Note that
\begin{equation}\label{eq:keydecomposition}
\begin{aligned}
  \nabla\widetilde{\mathcal{L}}(\widehat{\betada}) & \approx \nabla\widetilde{\mathcal{L}}({\betada}^{*}) +  \Jda (\widehat{\betada} -\betada^*) + (\nabla^2\widetilde{\mathcal{L}}(\widetilde{\betada}')-\Jda)(\widehat{\betada} -\betada^*)\\
  &=\underbrace{\nabla\widehat{\mathcal{L}}({\betada}^{*}) +  \Jda (\widehat{\betada} -\betada^*)}_{R_1} +  \underbrace{\nabla\widetilde{\mathcal{L}}({\betada}^{*})-\nabla\widehat{\mathcal{L}}({\betada}^{*})}_{R_2} + \underbrace{(\nabla^2\widetilde{\mathcal{L}}(\widetilde{\betada}')-\Jda)(\widehat{\betada} -\betada^*)}_{R_3},
\end{aligned}
\end{equation}
where $\widetilde{\betada}'$ is between $\widehat{\betada}$ and $\betada^*$. As long as conditions, (a) $\nabla\widetilde{\mathcal{L}}(\betada^*)\approx\nabla\widehat{\mathcal{L}}(\betada^*)$, (b) $\nabla^2\widetilde{\mathcal{L}}(\betada^*)\approx \Jda$, and (c) $\widehat{\betada}$ is sufficiently close to $\betada^*$, hold, we claim that $R_2$ and $R_3$ are some negligible small terms. Hence, combined with the KKT condition that $\nabla\widetilde{\mathcal{L}}(\widehat{\betada}) + \obC\widehat{\alphada} = \zeroda_p$ for some $\widehat{\alphada}\in\mathbb{R}^r$ and the constraint $\obC^\trans\widehat{\betada}=\tda$, similar derivations as before give that  $\nabla\widehat{\mathcal{L}}(\widehat{\betada})$ shares the same Kiefer-Bahadur representation as $\nabla\widehat{\mathcal{L}}(\widehat{\betada}^{\text{global}})$ up to some higher-order terms. This finding enables the use of $\nabla\widehat{\mathcal{L}}(\widehat{\betada})$ to construct a score-type test statistic for the linear hypothesis \eqref{eq:linearH0}. 

The above analysis motivates us to construct a surrogate loss $\widetilde{\mathcal{L}}(\betada)$ in a collaborative manner {that satisfies conditions~(a)--(c).} 
Without loss of generality, we treat the first data site with index set $\mathcal{I}_1$ as the master within the cluster, which can communicate with all other sites. A naive approximation of $\widehat{\mathcal{L}}(\betada)$ uses the local empirical loss on the master with a gradient shift:
\begin{equation}\label{eq:naivesurrgoate}
\widetilde{\mathcal{L}}^{\text{naive}}(\betada) = \widehat{\mathcal{L}}_1(\betada) + \langle \nabla\widehat{\mathcal{L}}(\betada^*) - \nabla\widehat{\mathcal{L}}_1(\betada^*), \betada  \rangle.
\end{equation}
It is easy to see that $\nabla\widetilde{\mathcal{L}}^{\text{naive}}(\betada^*)=\nabla\widehat{\mathcal{L}}(\betada^*)$ and $\nabla^2\widetilde{\mathcal{L}}^{\text{naive}}(\betada^*)=\nabla^2\widehat{\mathcal{L}}_1(\betada^*)$. We note that the Hessian shift $\frac{1}{2}\langle (\nabla^2 \hmL (\betada^*)-\nabla^2 \hmL_1 (\betada^*)  ) (\betada-\betada^*),\betada-\betada^*\rangle$ could be also added into the naive approximation \eqref{eq:naivesurrgoate} to guarantee that $\nabla^2\widetilde{\mathcal{L}}^{\text{naive}}(\betada^*)=\nabla^2\widehat{\mathcal{L}}(\betada^*)$, which approximates $\Jda=\nabla^2\mathcal{L}(\betada^*)$ more efficiently. However, this would require each site to transmit a $p\times p$ Hessian matrix $\nabla^2\hmL_k(\betada^*)$ to the master, leading to extra communicational burden. On the other hand, the gradient shift requires only a communication budget of $O(mp)$ bits within the cluster for transmitting the local gradients. In fact, given that $R_3$ in \eqref{eq:keydecomposition} is the product of the Hessian approximation error $\nabla^2\widetilde{\mathcal{L}}(\betada^*)-\Jda$ and the estimation error $\widehat{\betada}-\betada^*$, the  local Hessian approximation $\nabla^2\widehat{\mathcal{L}}_1(\betada^*)$ suffices to ensure that $R_3$ is negligible under certain scaling conditions on the local sample size $n_1=|\mathcal{I}_1|$. In addition, we allow considerable flexibility in the local sample sizes $n_k$ for all $k\neq1$, which can be arbitrarily small. 

The main issue with $\widetilde{\mathcal{L}}^{\text{naive}}(\betada)$ is that the truth $\betada^*$ is inaccessible. To address this, we  approximate $\betada^*$ using some initial estimator $\widehat{\betada}^{(0)}$, and define the surrogate collaborative loss as
\begin{equation}\label{eq:one-step}
\widetilde{\mathcal{L}}(\betada;\widehat{\betada}^{(0)}) = \widehat{\mathcal{L}}_{1}(\betada) + \langle  \nabla \widehat{\mathcal{L}}(\widehat{\betada}^{(0)}) - \nabla \widehat{\mathcal{L}}_{1}(\widehat{\betada}^{(0)}), \betada \rangle.
\end{equation}
Interestingly, we note that this formulation coincides with the gradient-enhanced loss proposed in \cite{wang2017efficient}, \cite{jordan2018communication} and \cite{fan2021communication}. Their heuristic motivation is that $\widetilde{\mathcal{L}}$ approximates the first-order Taylor remainder of the global empirical loss with the corresponding local Taylor remainder on the master, and it can improve the efficiency of an initial estimator. However, our construction of $\widetilde{\mathcal{L}}$ serves to approximate the critical components in the Kiefer-Bahadur representation for the traditional score statistic, which also offers a novel perspective on the gradient-enhanced loss. Hereafter, we use $\nabla\widetilde{\mathcal{L}}(\betada;\widehat{\betada}^\iternum{0})$ to denote the gradient of $\widetilde{\mathcal{L}}$ with respect to its first argument $\betada$.

\subsection{Two-stage partial penalization strategy}
 
In the previous discussions, we ignore the high-dimensional nuisance parameter $\bgamma$, which could significantly bias the Kiefer-Bahadur representation of the score statistic. In the following, we introduce a two-stage partial penalization strategy to both shrink the
approximation error of the surrogate loss (reducing $R_2$ and $R_3$ in \eqref{eq:keydecomposition}) and mitigate biases from high-dimensional components. We summarize the procedure in Algorithm~\ref{algo:est}. In Stage I, we use the standard $\ell_1$ penalization to obtain a relatively accurate estimator: 
\begin{equation}
\widehat{\betada}^\iternum{\mathrm{I}, k}\in\argmin_{\betada=(\btheta^\trans,\bgamma^\trans)^\trans }\widetilde{\mathcal{L}}(\betada;\widehat{\betada}^\iternum{\mathrm{I}, k-1})+\lambda_{\mathrm{I}, k}\normone{\bgamma},\quad k=1, \ldots, K_{\mathrm{I}}. \label{QR_dist_l1_0}
\end{equation}
In Stage II, starting from $\widehat{\betada}^\iternum{\mathrm{II}, 0}=\widehat{\betada}^{(\mathrm{I}, K_{\mathrm{I}})}$, a reweighted $\ell_1$ penalization combined with linear constraints is used for $k = 1, \ldots, K_{\mathrm{II}}$, in order to alleviate the non-negligible bias due to the $\ell_1$ penalty in the first stage: 
\begin{equation} 
 \widehat{\betada}^\iternum{\mathrm{II}, k}\in\argmin_{\betada=(\btheta^\trans,\bgamma^\trans)^\trans\atop\mathbf{C}\btheta=\tda}\widetilde{\mathcal{L}}(\betada;\widehat{\betada}^\iternum{\mathrm{II}, k-1})+\sum_{j=1}^{p-d} q_{\lambda_{\mathrm{II}, k}}'(|\widehat{\gamma}_{j}^{(\mathrm{II}, k-1)}|)|\gamma_j|, \label{eq:stage2eq}
\end{equation}
where $q_\lambda(\cdot)$ is some folded-concave penalty specified later.

In the end, we make some comments regarding the two stage partial penalization strategy. First, to shrink the
approximation error of $R_2=\nabla\widetilde{\mathcal{L}}(\betada^*;\betada)-\nabla\hmL(\betada^*)$, it is necessary to reduce the initialization error, i.e. the distance between $\betada$ and $\betada^*$.  The first stage is designed to provide a reasonably accurate estimator, which is crucial to ensure the reliability of the reweighted regularization parameter in Stage II. The second stage then refines this estimator by mitigating the bias introduced in Stage I via a folded-concave penalty, facilitating tractable asymptotic behavior of our test statistic. Although a constrained formulation could also be used in Stage I, imposing such constraints under the alternative hypothesis may introduce non-negligible bias, undermining the theoretical guarantees of the Stage II estimator. Second, the weighted regularization in \eqref{eq:stage2eq} operates under a collaborative setting with linear constraints, which is significantly different from the unconstrained shared-data setting \citep{zou2008one,fan2014strong} and requires novel technical developments. Third, partial penalization guarantees that no penalties are imposed on $\btheta$, the parameters of interest in our hypothesis. This can avoid undesirable bias induced by the sparsity penalty on the final estimator of $\btheta^*$ \citep{shi2019linear}. Moreover, it circumvents the need to impose a minimal signal condition on $\btheta^*$, thereby ensuring that the corresponding test retains adequate power even at local alternatives. Our last remark regards the communication complexity.  Algorithm~\ref{algo:est} needs a communication budget of $O(mp)$ bits within the cluster per iteration (for transmitting the current estimator and local gradients). This is considered communication-efficient so long as the number of iterations remains reasonable \citep{fan2021communication}. The requirement for the iteration numbers will be evaluated in Section~\ref{sec:theory}.

\begin{algorithm}[ht]
 \caption{Two-stage partial penalized collaborative estimation} 
 \label{algo:est}
 \begin{algorithmic}[1] 
  \setlength{\baselineskip}{10pt} 
 \Require Initialization {
 $\widehat{\betada}^{(0)}$}, maximum step $K_{\mathrm{I}}$ and $K_{\mathrm{II}}$. 
 \For{$t = \mathrm{I}, \mathrm{II}$} 
 \State Set the initial estimator as the final estimator in the last stage.
 \For{$k = 1, \ldots, K_t$} 
 \State The master broadcasts the current estimator to all local sites. Each site evaluates its local gradient and sends it to the master.
 \State The master aggregates the global gradient, and then solves \eqref{QR_dist_l1_0} when $t=\mathrm{I}$, and \eqref{eq:stage2eq} when $t=\mathrm{II}$ to update the estimator.
 \EndFor 
 \EndFor
 \Ensure{The final estimator $\widehat{\betada}=\widehat{\betada}^{(\mathrm{II}, K_{\mathrm{II}})}$.}
 \end{algorithmic} 
\end{algorithm} 

\subsection{Construction of the CST}
We collect some general notations. Let $\mathcal{S}=\{j\in[p-d]:\gamma^*_j\neq0\}$ be the support of $\bgamma^*$ and the sparsity level $s=|\Sgamma|$. For any $\mathcal{T}\subseteq[p-d]$, we adopt the convention that $ {\mathcal{T}_a}=[d]\cup\{j+d:j\in\mathcal{T}\}$ for convenience. Denote $\boldsymbol{b}$ as the subvector $\betada_{\os}=(\btheta^\trans,\bgamma_\Sgamma^\trans)^\trans$ of $\betada$. Hence, we have $\nablabb =  (\nabla_{\btheta}^\trans,\nabla_{\bgamma_{\Sgamma}}^\trans)^\trans$ and $\widehat{\bda} = (\widehat{\btheta}^\trans,\widehat{\bgamma}_{\Sgamma}^\trans)^\trans$. Let  $\Jda_0 = \nablabb^2\mathL(\betada^*)$ and $\bK_0 = \mathbb{E}[\nabla_\bda\ell(\betada^*;\zda_i)(\nabla_\bda\ell(\betada^*;\zda_i))^\trans]$ represent the Hessian of the risk function and the covariance of the score function with support $\os$, respectively. Note that when $\ell(\cdot)$ is the negative log-likelihood function, the second Bartlett identity says that $\bK_0=\Jda_0$. Nevertheless, this identity may not be true for a general loss, which we consider in this article. Further define the corresponding sub-matrices for $\bC_a$, $\Psi$ and $\mathbf{P}$: $\bC_{a0}=(\bC,\mathbf{O}_{r\times s})^\trans\in\mathbb{R}^{(d+s)\times r}$, $\Psi_0 = \bC_{a0}^\trans\Jda_0^{-1}\bC_{a0}$, and $\mathbf{P}_0 =\Jda_0^{-1/2}\bC_{a0}\Psi_0^{-1}\bC_{a0}^\trans\Jda_0^{-1/2}$. 

Before constructing the CST, we first introduce an oracle collaborative score-type test (OCST), which assumes prior knowledge of the true support of the nuisance parameter $\bgamma^*$.  For $k=1,\ldots,K_{\mathrm{II}}$, define the oracle version of the estimators in Stage II:
\begin{equation}\label{eq:disora_def}
 \widehat{\betada}^{\ora,(k)}=\argmin_{\bC\btheta=\tda,\,\bgamma_{\Sgammac}=\zeroda}\widetilde{\mathcal{L}}(\betada;\widehat{\betada}^\iternum{\mathrm{II},k-1}).
\end{equation}
Let $\widehat{\betada}^\ora=\widehat{\betada}^{\ora,(K_{\mathrm{II}})}$. Since $\widehat{\betada}^{\ora}$ is essentially a low-dimensional estimator based on the surrogate loss. Following the heuristic discussions in Section~\ref{subsec:decomposition}, we have the asymptotic Kiefer-Bahadur representation for the score statistic:
$$
\nabla_{\bda}\hmL(\widehat{\betada}^\ora) \approx \bC_{a0}\Psi_0^{-1} \left(\bC_{a0}^\trans\Jda_0^{-1}\nabla_\bda\hmL(\betada^*)-\hda\right).
$$
Therefore, it motivates us to construct the following OCST statistic:
\[
T^\ora_S = N\Vert{\boldsymbol{\Omega}}\nabla_{\bda}\hmL(\widehat{\betada}^\ora)\Vert_2^2,
\]
where $\boldsymbol{\Omega} = \bV ^{-1/2}\bC_{a0}^\trans\Jda_0^{-1}$ and $\bV = \bC_{a0}^\trans \Jda_0^{-1}\bK_0\Jda_0^{-1}\bC_{a0}$. Under appropriate conditions, we can show that the asymptotic distribution of $T^\ora_S$ is a $\chi^2$ distribution with $r$ degrees of freedom under $H_0$.

Although the OCST statistic $T^\ora_S$ is inaccessible in practice due to the lack of prior knowledge on $\Sgamma$, we claim that the two-stage strategy can effectively estimate $\Sgamma$ with high probability. Let $\widehat{\betada}=\widehat{\betada}^\iternum{\mathrm{II},K_{\mathrm{II}}}$, and choose $\widehat{\Jda}_0$ and $\widehat{\bK}_0$ as any estimators of $\Jda_0$ and $\bK_0$, which also rely on the support estimator $\widehat{\Sgamma}= {\mathrm{supp}(\widehat{\bgamma})}$. Let $\widehat{\bC}_{a0}=(\bC,\bO_{r\times |\widehat{\Sgamma}|})^\trans\in\mathbb{R}^{(d+|\widehat{\Sgamma}|)\times r}$. Then we have the corresponding estimators $\widehat{\bV}=\widehat{\bC}_{a0}^\trans (\widehat{\Jda}_0)^{-1}\widehat{\bK}_0(\widehat{\Jda}_0)^{-1}\widehat{\bC}_{a0}$ and $\widehat{\boldsymbol{\Omega}} = \widehat{\bV}^{-1/2}\widehat{\bC}_{a0}^\trans(\widehat{\Jda}_0)^{-1}$ for $\bV$ and $\boldsymbol{\Omega}$, respectively. We propose the CST statistic as
\begin{equation}\label{eq:CST_statistic}
T_S = N\Vert\widehat{\boldsymbol{\Omega}}\nabla_{\widehat{\bda}}\hmL(\widehat{\betada})\Vert_2^2.
\end{equation}
Our theoretical results  show that under mild conditions, the CST exhibits an oracle property in the sense that  $T_S$ shares the same limiting distribution as the OCST statistic $T_S^\ora$, provided $\widehat{\boldsymbol{\Omega}}$ is consistent. Thus we can use $T_S$ to test the linear hypothesis \eqref{eq:linearH0}: for a given significance level $\alpha$, we reject $H_0$ when $T_S>\chi^2_\alpha(r)$, where $\chi^2_\alpha(r)$ is the upper $\alpha$ quantile of $\chi^2$ distribution with $r$ degrees of freedom. In the next section, we will validate the CST procedure theoretically. 

The remaining problem is to provide a consistent estimate for $\boldsymbol{\Omega}$. When the communication budget allows and the second Bartlett identity fails, one can directly employ the empirical Hessian $\widehat{\Jda}_0=\sum_{i=1}^N \nabla_{\widehat{\bda}}^2\ell(\widehat{\betada};\zda_i)/N$ and covariance of score function $\widehat{\bK}_0=\sum_{i=1}^N \nabla_{\widehat{\bda}}\ell(\widehat{\betada};\zda_i)\nabla_{\widehat{\bda}}\ell(\widehat{\betada};\zda_i)^\trans/N$ using the full samples. However, we need to transmit the large scale matrices from each local site to the master, which might lead to excessive communication costs. To achieve a trade-off between communication efficiency and statistical accuracy, we suggest to use an averaged test statistics with local variance estimators for $\bV$ and $\boldsymbol{\Omega}$ when the sample size of each data source is balanced. Specifically, let $\widehat{\boldsymbol{\Omega}} =\sum_{k=1}^m n_k \widehat{\boldsymbol{\Omega}}_k/N$, where $\widehat{\boldsymbol{\Omega}}_k$ are constructed based on the local variance estimators $\widehat{\Jda}_{0,k}$ and $\widehat{\bK}_{0,k}$ at the $k$th site. Since $T_{S} = \sum_{k=1}^m n_k\Vert\widehat{\boldsymbol{\Omega}}_k\nabla_{\widehat{\bda}}\hmL(\widehat{\betada})\Vert_2^2$ in this case, constructing the CST statistic only requires transmitting the gradient $\nabla_{\widehat{\bda}} \hmL(\widehat{\betada})$ to each data site, which then returns the one-dimensional quantity $n_k \Vert \widehat{\boldsymbol{\Omega}}_k \nabla_{\widehat{\bda}} \hmL(\widehat{\betada}) \Vert_2^2$ to the master. If a data source has notably fewer samples compared to others, we could exclude it in the variance estimation to prevent extra bias due to the limited sample size.

\section{Statistical analysis}\label{sec:theory}
We begin by outlining the general assumptions for our theoretical derivations, followed by some key quantities relevant to the technical results. Then, we present the main theoretical findings on the CST procedure. Finally, we consider the GLM as an illustrative example.
\subsection{General assumptions} 
\begin{assump}[Folded-concave penalty]\label{assump:penalty}
 $q_\lambda(\cdot)$ is of the form $q_\lambda(t)=\lambda^2 q(t / \lambda)$ for $t \ge 0$, where $q:[0, \infty) \to [0, \infty)$ satisfies: (i) $q(\cdot)$ is non-decreasing and concave on $[0, \infty)$ with $q(0)=0$; (ii) $q'(\cdot)$ exists almost everywhere and is non-increasing on $(0, \infty)$. Moreover, $0 \le q'(t) \le 1$ and $\lim _{t \downarrow 0} q'(t)=1$; (iii) $q'(\alpha_1)=0$ for some $\alpha_1 > 0$.
\end{assump}
 
Examples satisfying Assumption~\ref{assump:penalty} include the smoothly clipped absolute deviation (SCAD) penalty \citep{fan2001variable} and the minimax concave penalty (MCP) \citep{zhang2010nearly}. The $\ell_1$ penalty \citep{tibshirani1996regression} satisfies Assumption~\ref{assump:penalty} without property (iii).

\begin{assump}[Local identifiability]\label{assump:regular}
 There exist $0<J_l\le J_u\lesssim1$ and $0<K_l\le K_u\lesssim1$ such that $J_l\le\lambda_{\min}(\Jda)\le\lambda_{\max}(\Jda)\le J_u$ and $K_l\le\lambda_{\min}(\bK_0)\le\lambda_{\max}(\bK_0)\le K_u$.
\end{assump}
Assumption~\ref{assump:regular} requires that both the score function $\nabla\mathcal{L}(\betada^*)$ and the Hessian $\nabla^2\mathcal{L}(\betada^*)$ are non-degenerate, which is also a local identifiability condition for the minimizer $\betada^*$.

\begin{assump}[Concentration of score]\label{assump:score}
 There exist $\eta_\infty(N)$ and $\eta_2(N)$ such that $\norminf{\nabla\hmL(\betada^*)} \le \eta_\infty(N)$, and $\normtwo{\nabla_{\bda}\hmL(\betada^*)}\le\eta_2(N)$.
\end{assump}  
Assumption~\ref{assump:score} essentially imposes a concentration condition on the empirical gradient with respect to certain norms since $\nabla\mathcal{L}(\betada^*) =\zeroda$. We refer to Assumption~3.2 in \citet{Ning2017} for more discussions on this assumption for high-dimensional inference.

\begin{assump}[Homogeneity]\label{assump:homo}
 There exist $\delta_\infty$ and $\delta_{2}\in (0,1)$ such that $\Vert\nabla^2\hmL(\betada^*)-\Jda\Vert_{\max} \vee \normmax{\nabla^2\hmL_1(\betada^*)-\Jda}\le \delta_\infty$, and $\Vert\nabla_{\bda}^2\hmL_1(\betada^*)-\Jda_0\Vert_{2}\le \delta_{2}$. 
\end{assump}
Assumption~\ref{assump:homo} introduces a homogeneity condition to control variations in loss functions across local data sites, akin to the conditions in \citet{fan2021communication} (see Assumption~3.2 therein). Additionally, we require that the Hessian matrix of the loss function on the master machine closely approximates its population counterpart, which essentially necessitates a scaling of local sample size for applying certain high probability bounds.

Before stating Assumptions~\ref{assump:lip} and \ref{assump:rsc}, we introduce some notation. Let $\kappa$ be a positive constant such that there exists some  constant $\alpha_0>0$ such that 
 \begin{equation}\label{eq:def_phialpha0}
 \phi:=\dfrac{\sqrt{1+[q'(\alpha_0)/2]^2}}{\alpha_0\kappa}\in(0,1) \quad \textrm{and} \quad q'(\alpha_0)>0.
 \end{equation}
 Let $\mathtt{g} = (2+2/q'(\alpha_0))(c^2+1)^{1/2} (d+s)^{1/2}$, where $c>0$ is defined through the equation
 \begin{equation}\label{eq:c_def}
 0.5q'(\alpha_0)(c^2+1)^{1/2} + 1 = \alpha_0\kappa c.
 \end{equation}
{For example, consider the MCP defined by $q'_{\mathrm{MCP}}(t) = (1 - t/a) \vee 0$ for $t \ge 0$ with $a > 1$. Given $\kappa>0$, if $a>4\kappa^{-1}$, then $\alpha_0 = a/2$ satisfies \eqref{eq:def_phialpha0}. In addition, equation \eqref{eq:c_def} admits a unique solution $c > 0$ since $0.5 q'(\alpha_0) < \alpha_0 \kappa$.}
 
\begin{assump}[Restricted Lipschitz Hessian]\label{assump:lip}
 Both the local and global empirical loss function $\hmL_1(\betada)$ and $\hmL(\betada)$ have restricted Lipschitz Hessian for some radius $R_{\mathrm{lip}}>0$. That is, for all $\vda \in \Lambda(\mathtt{g})\cap\Theta(R_{\mathrm{lip}})$, it holds that
 \begin{align*}
 \norminf{\{\nabla^2\hmL_1(\betada^*+\vda)-\nabla^2\hmL_1(\betada^*)\}\vda}&\le M \Vert{\vda}\Vert_2^2,\\
 \text{and}\quad \norminf{\{\nabla^2\hmL(\betada^*+\vda)-\nabla^2\hmL(\betada^*)\}\vda}&\le M \Vert{\vda}\Vert_2^2,
 \end{align*}
 where $M$ is some non-negative constant.
\end{assump}

\begin{assump}[Restricted local strongly convexity]\label{assump:rsc}
 The local loss function $\hmL_1(\betada)$ on the master machine is restricted $\kappa$-strongly convex for some radii $R_{\mathrm{loc}}$ and $R_{\mathrm{rsc}}>0$: for all $\betada=(\btheta^\trans,\bgamma^\trans)^\trans$ satisfying $\normtwo{\betada-\betada^*}\le R_{\mathrm{loc}}$ and $\bgamma_{\Sgammac}=\zeroda$, and $\vda \in \Lambda(\mathtt{g})\cap\Theta(R_{\mathrm{rsc}})$,
 \[
 \langle \nabla\hmL_1(\betada+\vda)-\nabla\hmL_1(\betada),\vda \rangle \ge \kappa\Vert\vda\Vert_2^2.
 \]
\end{assump}
Assumptions~\ref{assump:lip} and \ref{assump:rsc} concern the restricted Lipschitz Hessian matrix and restricted strongly convexity (RSC) of the empirical loss function, which are standard for high-dimensional problems \citep{loh2017support,fan2018lamm,jordan2018communication}.

\begin{assump}\label{assump:h_end}
 Assume that $\normtwo{\hda}\in[0,\sqrt{r/N}]$ satisfies $\normtwo{\hda}\lesssim  R_{\mathrm{loc}} \wedge \eta_\infty(N)\wedge \eta_2(N)$, and $0<\lambda_{\min}(\Cda\Cda^\trans)\le\lambda_{\max}(\Cda\Cda^\trans) \lesssim 1$.
\end{assump}

Assumption~\ref{assump:h_end}  focuses on the local alternative parameter $\hda = \Cda\btheta^* - \tda$, where the sequence of vectors $\hda\to\zeroda$ as $N\to\infty$. It limits our consideration to local alternatives with a bounded radius for $\normtwo{\hda}$ and also requires the regularity of the constraint matrix $\bC$. {Typical examples include $ \bC_{\text{id}} = ( \mathbf{I}_{r \times r},\mathbf{O}_{r \times (d-r)})$ and the first-order difference matrix $\bC_{\text{diff}} \in \mathbb{R}^{ r\times d}$, where the $(i,i)$th entry is $1$, the $(i,i+1)$th entry is $-1$, and all others are $0$.} 

\subsection{Technical quantities}
For any $k\ge1$, define the estimation error for stage I as $r_{\mathrm{I},k}=\normtwo{\widehat{\betada}^{(\mathrm{I},k)}-\betada^*}$.  In particular, the initialization error is denoted by $r_0=\normtwo{\widehat{\betada}^{(\mathrm{I},0)}-\betada^*}$. We set the regularization parameters in \eqref{QR_dist_l1_0} and \eqref{eq:stage2eq} as
 \begin{gather}
 \lambda_{\mathrm{I}, k} = C_1\left( \delta_\infty(d+s)^{1/2}r_{\mathrm{I},k-1} + Mr_{\mathrm{I},k-1}^2 +\eta_\infty(N)\right) \label{eq:lambda1} \\
\text{and} \,\,\,\,\, \lambda_{\mathrm{II}, k}\equiv\lambda_{\mathrm{II}} = C_2\left( \eta_\infty(N) + M \normtwo{\hda}^2 + d^{1/2}\delta_\infty\normtwo{\hda}\right)\quad  \label{eq:lambda2}
 \end{gather}
for some sufficiently large constants $C_1$ and $C_2$, which are used to control the magnitude of random noise. Further define the optimal errors for each stage: 
\begin{gather}
        r^*_{\mathrm{I}} = \alpha_0 c(d+s)^{1/2}\lambda_{\mathrm{II}}  \quad \text{and} \\
      r^*_{\mathrm{II}}=\eta_2(N) + M (d+s)^{1/2} \normtwo{\hda}^2 + (d+s)\delta_\infty\normtwo{\hda} + \normtwo{\hda}.
\end{gather}
In addition, define the contraction factors for each stage: $f_{\mathrm{I}}=\mathtt{f}(r_0)$ and $f_{\mathrm{II}}=\mathtt{f}(r^*_{\mathrm{I}})$, where $\mathtt{f}(r)\asymp (d+s)\delta_\infty + M(d+s)^{1/2}r$. Further let the optimal contraction factor be 
$f^*_{\mathrm{II}} = \mathtt{f}(r^*_{\mathrm{II}})$. {We provide some remarks on the optimal errors and contraction factors below.}
\begin{rmk}[Optimal error]
    The optimal Stage I error $r^*_{\mathrm{I}}$ has a similar form to the $\ell_1$-penalized estimation method \citep{negahban2012unified}. The optimal Stage II error $r^*_{\mathrm{II}}$ consists of two parts: $\eta_2(N)$ represents the intrinsic optimal error of the estimation problem, and the terms involving $\normtwo{\hda}$ capture the bias induced by the alternative hypothesis for the constrained estimator.  
\end{rmk}
\begin{rmk}[Contraction factor]
    The contraction factors comprise two main components. First, the homogeneity parameter $\delta_\infty$ controls the variability of gradient shifts. Second, the accuracy of the initial estimator is crucial for ensuring contraction for each iteration.  In Appendix~A of the Supplementary Material, we elucidates that the multi-iteration procedure refines the statistical error in a geometric rate. Specifically, in each iteration, the estimation error reduces by a factor of $f_{\mathrm{I}}$ or $\overline{f}_{\mathrm{II}}=(f_{\mathrm{II}} + \phi)$ for each stage, leaving residual error terms of order $r^*_{\mathrm{I}}$ and $r^*_{\mathrm{II}}$, respectively. 
\end{rmk}

\subsection{Theoretical results}
Now, we examine the limiting distribution of the CST statistic $T_S$ under both the null hypothesis and local alternatives. Before that, we first derive the limiting distribution of the OCST statistic $T_S^\ora$.  Our analysis is based on two main components: the Kiefer-Bahadur representation remainder decays at a certain rate, and a Lyapunov-type condition is satisfied to ensure the asymptotic normality. We collect the required assumptions in Condition~\ref{cond:general}.

\begin{condition}\label{cond:general}
  Suppose that (i) Assumptions~\ref{assump:penalty}--\ref{assump:h_end} hold with  $R_{\mathrm{lip}}>r_{0}\vee r_{\mathrm{I}}^{*}$  and 
 $R_{\mathrm{rsc}} > (1.5(d+s)^{1/2} \kappa^{-1}\lambda_{\mathrm{I},1})\vee r_{\mathrm{I}}^{*}$; (ii) the initial estimator satisfies $\widehat{\betada}^{(\mathrm{I},0)}-\betada^*\in\Lambda(\mathtt{g})$;  (iii) the beta-min condition $|\bgamma^*_{\Sgamma}|\ge(\alpha_0+\alpha_1)\lambda_{\mathrm{II}}$ holds for the nuisance parameter; (iv)  $\mathbb{E}\normtwo{\boldsymbol{\Omega} \nabla_\bda\ell(\betada^*;\zda_i)}^3 = o(N^{1/2})$;   (v) the contraction factors $f_{\mathrm{I}}<1$ and $\overline{f}_{\mathrm{II}} = f_{\mathrm{II}} +\phi<1$; (vi) $f^*_{\mathrm{II}} r^*_{\mathrm{II}}  = o(N^{-1/2})$; and (vii) the iteration numbers $K_{\mathrm{I}}\gtrsim \log(r_0/r^*_{\mathrm{I}})/\log(1/f_{\mathrm{I}})$ and $K_{\mathrm{II}}\gtrsim \log(r_{\mathrm{I}}^{*}/r^*_{\mathrm{II}})/\log(1/\overline{f}_{\mathrm{II}})$.
\end{condition}
\begin{rmk}
  Condition~(ii) requires that the initial estimator lies in an $\ell_1$ cone centered at $\betada^*$, which is a standard result for $\ell_1$-penalized estimators \citep{bickel2009simultaneous,negahban2012unified}. For example, we can use the local $\ell_1$-penalized estimator computed on the master machine.  Condition~(iii) imposes a minimal signal condition on the nuisance parameter, which is mild in high-dimensional problems \citep{fan2014strong,loh2017support}. In particular, we do not need  a similar condition on the target parameter $\btheta^*$, allowing the test to  retain adequate power even at local alternatives. Condition~(iv) is a Lyapunov-type condition that ensures the asymptotic normality. Condition~(v) assumes that the contraction factors are strictly less than $1$, which guarantees that we can shrinkage the
  approximation error of the surrogate loss and mitigate biases from high-dimensional components in each iteration.  Condition~(vi) ensures that the Kiefer-Bahadur representation remainder for $T_S^\ora$ decays faster than $N^{-1/2}$.  Condition~(vii) leverages the geometric convergence of the two-stage strategy: the iteration numbers $K_{\mathrm{I}}$ and $K_{\mathrm{II}}$ are chosen to ensure that the estimators achieve the optimal error rate at each stage.
\end{rmk}

Theorem~\ref{thm:oraCST} establishes the limiting distribution of the OCST statistic. We note that the limiting distribution still holds even when the number of constraints $r$ and the number of the target parameters $d$ diverge with the sample size at an appropriate rate. 
\begin{thm}\label{thm:oraCST}
 Under Condition~\ref{cond:general},  we have
 \[
 \sup_{x\ge0}|\mathbb{P}(T^\ora_S\le x)-\mathbb{P}(\chi^2(r,N\hda^\trans \bV^{-1}\hda) \le x)|\to 0.
 \]
\end{thm}

Next, we show that the CST statistic enjoys an oracle property that its limiting behavior is the same as that of the OCST statistic, despite the agnostic support for $\bgamma^*$. To this end, we introduce some additional assumptions in Condition~\ref{cond:general2}.

\begin{condition}\label{cond:general2}
  In addition to Condition~\ref{cond:general}, suppose that: (i) $f_{\mathrm{II}} r_{\mathrm{I}}^{*}\lesssim \lambda_{\mathrm{II}}$;  (ii) 
 $ \norminf{\Jda_0^{-1/2}(\mathbf{I}-\mathbf{P}_0)\Jda_0^{-1/2}\nabla_\bda\hmL(\betada^*)}\vee \norminf{\Jda_0^{1/2}(\mathbf{I}-\mathbf{P}_0)\Jda_0^{-1/2}\nabla_\bda\hmL(\betada^*)}\le\eta_\infty(N)$ in Assumption~\ref{assump:score}; (iii)   $R_{\mathrm{loc}} > r_{\mathrm{I}}^{*}$ in Assumption~\ref{assump:rsc};  (iv) there exists {$A_0\in(0,\infty)$} such that $\max_{j\in\osc} \normone{\Jda_{j,\os}(\Jda_0)^{-1}}\le A_0$; (v) $\normtwo{\widehat{\Jda}_0-\Jda_0}\vee\normtwo{\widehat{\bK}_0-\bK_0}=o_p(1)$; and (vi) $K_{\mathrm{II}}\gtrsim \lceil\log(\widehat{s})\rceil$, where $\widehat{s}=|\mathrm{supp}(\widehat{\bgamma}^\iternum{\mathrm{II},0})|+s$.
\end{condition}

\begin{rmk}
Conditions~(i) and (iii) are easily satisfied with some scaling conditions on the sample size. Condition~(ii) holds when the random noise $\nabla\widehat{\mathcal{L}}(\betada^*)$ has a light tailed distribution under Assumption~\ref{assump:score}. Notably,  Condition~(iv) is significantly weaker than the irrepresentable condition, which is both sufficient and nearly necessary for the model consistency of the Lasso \citep{zhao2006model}. A common form of the irrepresentable condition can be formulated as follows: for a given value $a_0\in(0, 1)$, $\max_{j\in\mathcal{S}^c}\normone{\Jda_{j\mathcal{A}}\Jda_{\mathcal{A}\mathcal{A}}^{-1}}\le a_0$, where $\mathcal{A}$ is the true support of the specific problem. Condition~(v) ensures the consistency of the variance estimator.  Finally,  the additional requirement for $K_{\mathrm{II}}$  in Condition~(vi) determines when the CST statistic achieves the oracle property. Further details on the iteration number and the oracle property are provided in Appendix~C of the Supplementary Material.
\end{rmk}

\begin{thm}\label{thm:CST}
 Under Condition~\ref{cond:general2}, we have
 \[
 \sup_{x\ge0}|\mathbb{P}(T_S\le x)-\mathbb{P}(\chi^2(r,N\hda^\trans \bV^{-1}\hda) \le x)|\to 0.
 \]
\end{thm}

Recall that the CST is conducted as follows: for a given significance level $\alpha$, we reject the null hypothesis when $T_S>\chi^2_\alpha(r)$. Corollary~\ref{cor:test} validates the testing procedure.

\begin{cor}\label{cor:test}
Suppose the conditions in Theorem~\ref{thm:CST} hold. Under the null hypothesis, for any $0<\alpha<1$, we have 
 \[
 \lim \mathbb{P}(T_S>\chi^2_\alpha(r)) = \alpha.
 \]
Under the local alternative with $\hda=\bC\btheta^*-\tda$ satisfying Assumption~\ref{assump:h_end}, we have 
 \[
 \lim|\mathbb{P}(T_S > \chi^2_\alpha(r))-\mathbb{P}(\chi^2(r,e_N) > \chi^2_\alpha(r))|=0.
 \]
\end{cor}
\begin{rmk}
  Corollary~\ref{cor:test} shows that the asymptotic power function of the CST is $\mathbb{P}(\chi^2(r,e_N)>\chi^2_\alpha(r))$, where $e_N=N\hda^\trans \bV^{-1}\hda$. This result aligns with the power functions of traditional score tests in low-dimensional scenarios with shared data \citep{boos1992score}, demonstrating the efficiency gains from collaboration over independent local tests. Moreover, the CST is potentially more powerful than the DC test proposed by \cite{battey2018distributed}, since it circumvents the debiasing procedure.  For a numerical comparison, we refer to Section~\ref{subsec:dc}, with further discussions provided in {Appendix~E} of the Supplementary Material.
\end{rmk}

\subsection{Example}\label{sec:example}

We consider the GLM with canonical link as a demonstrating example. Suppose that $\xda_i=(\vda_i^\trans,\boldsymbol{q}_i^\trans)^\trans\in\mathbb{R}^p$ and the response variable $y_i$ has the conditional density function
\[
h(y_i|\xda_i,\betada^*) = c(\xda_i,y_i)  \exp\left\{ y_i( \vda_i^\trans\btheta^* + \boldsymbol{q}_i^\trans\bgamma^*) - b( \vda_i^\trans\btheta^*+ \boldsymbol{q}_i^\trans\bgamma^*)  \right\},
\]
where $\vda_i\in\mathbb{R}^d$, $\boldsymbol{q}_i\in\mathbb{R}^{p-d}$, $b(\cdot)$ is some known convex function, and $c(\cdot)$ is a known function such that $h$ is a valid probability density function. For simplicity, we let the dispersion parameter to be one as we do not consider the issue of over-dispersion.

In this case, we have $\zda_i = (y_i,\xda_i^\trans)^\trans\in\mathbb{R}^{p+1}$ and $\ell(\betada;\zda_i) = b(\xda_i^\trans\betada)-y_i(\xda_i^\trans\betada)$. Assume that $\bgamma^*$ has the strong sparse structure with support $\Sgamma$ and sparsity level $s=|\Sgamma|$. Let $\Sigma=\mathbb{E}[\xda_i\xda_i^\trans]$, and direct calculations give that  $
\nabla\ell(\betada;\zda_i)=[b'(\xda_i^\trans\betada)-y_i]\xda_i$ and $
\Jda = \bK = \mathbb{E}[b''(\xda_i^\trans\betada^*)\xda_i\xda_i^\trans]$. We invoke some standard regularity assumptions on the GLM.

\begin{assump} \label{cond:glm}
\begin{itemize}
    \item[(i)] There exist some positive constants $\sigma_l,\sigma_u, J_l$ and $J_u$ such that  $0<\sigma_l<\lambda_{\min}(\Sigma)\le\lambda_{\max}(\Sigma)<\sigma_u<\infty$ and $J_l<\lambda_{\min}(\Jda)\le\lambda_{\max}(\Jda)<J_u $;  
    \item[(ii)] The covariate $\xda_i$ is a $\sigma_x$-sub-Gaussian vector with bounded coordinates $\norminf{\xda_i}\le b_1$, and the random error $\nabla\ell(\betada^*;\zda_i)=[b'(\xda_i^\trans\betada^*)-y_i]\xda_i$ is a $\sigma_e$-sub-Gaussian vector;  that is, for all $\alphada\in\mathbb{R}^p$, $\mathbb{E}[\exp(\alphada^\trans\xda_i)]\le\exp(\sigma_x^2\normtwo{\alphada}^2/2)$ and $\mathbb{E}[\exp(\alphada^\trans \nabla\ell(\betada^*;\zda_i))]\le\exp(\sigma_e^2 \normtwo{\alphada}^2/2)$;
    \item[(iii)]  $|b''(\cdot)|$ and $|b'''(\cdot)|$ are bounded by some constants $b_2$ and $b_3$, respectively; 
    \item[(iv)] There exist positive constants $\kappa$ in \eqref{eq:def_phialpha0} and $L$ such that for any $\betada$ satisfying $\normtwo{\betada-\betada^*}\le R_{\mathrm{loc}}$ and $\vda\in\Lambda(\mathtt{g})\cap\Theta(R_{\mathrm{rst}})$ with some radii $R_{\mathrm{loc}}$ and $R_{\mathrm{rst}}$, we have $\langle\nabla\widehat{\mathcal{L}}_1(\betada+\vda)-\nabla\widehat{\mathcal{L}}_1(\betada),\vda\rangle\ge\kappa\normtwo{\vda}^2$ and $[\frac{1}{n_1}\sum_{i\in\mathcal{I}_1}(\xda_i^\trans\vda)^2 ] \vee [\frac{1}{N}\sum_{i\in[N]}(\xda_i^\trans\vda)^2 ]\le L \normtwo{\vda}^2$;
    \item[(v)] There exists $A_0\in(0,\infty)$ such that $\max_{j\in\osc} \normone{\Jda_{j,\os}(\Jda_0)^{-1}}\le A_0$; 
\end{itemize}   
\end{assump}
Assumption~\ref{cond:glm} requires sub-Gaussian covariates with positive definite covariance matrices. The positive definiteness of the Hessian essentially ensures that $b''(\xda_i^\trans\betada^*)$ is bounded away from zero with high probability, which is mild for sub-Gaussian designs. We also assume the random noise $\nabla\ell(\betada^*;\zda_i)$ to be sub-Gaussian, which is typical in high-dimensional analysis. These assumptions are common in the study of the GLMs \citep{negahban2012unified,fan2021communication}, and hold for linear, logistic, and multinomial models. Moreover, Assumption (v) is considerably weaker than the standard irrepresentable condition \citep{zhao2006model}, as it only requires $A_0$ to be any positive constant.

The following proposition verifies Assumptions~\ref{assump:regular}--\ref{assump:rsc} for the GLM.
 \begin{prop}\label{prop:glm_assump}
  For any $t_N>0$, under Assumption~\ref{cond:glm} and the sample scaling $n_1\gtrsim (d+s)\log(p)+t_N$, the following results hold for the GLM with probability at least $1-\exp(-t_N)$:
 \begin{enumerate}
  \item[(a)] \textbf{Concentration bounds}:
  \begin{itemize}
      \item[(i)] $\norminf{\nabla\hmL(\betada^*)} \le \eta_\infty(N)\asymp\sqrt{\{\log(p)+t_N\}/N}$; 
      \item[(ii)] $\normtwo{\nabla_{\bda}\hmL(\betada^*)}\le\eta_2(N) \asymp\sqrt{\{(d+s)+t_N\}/N}$;
      \item[(iii)] $\Vert\nabla^2\hmL(\betada^*)-\Jda\Vert_{\max} \vee \normmax{\nabla^2\hmL_1(\betada^*)-\Jda}\le \delta_\infty\asymp \sqrt{\{\log(p)+t_N\}/n_1}$;
      \item[(iv)]  and $\Vert\nabla_{\bda}^2\hmL_1(\betada^*)-\Jda_0\Vert_{2}\le \delta_{2}\asymp\sqrt{\{(d+s)+t_N\}/n_1}$.
  \end{itemize} 
 \item[(b)] \textbf{Restricted Lipschitz Hessian}: for all $\vda \in \Lambda(\mathtt{g})\cap\Theta(R_{\mathrm{rst}})$,
 \begin{align*}
 \norminf{\{\nabla^2\hmL_1(\betada^*+\vda)-\nabla^2\hmL_1(\betada^*)\}\vda}&\le M \Vert{\vda}\Vert_2^2,\\
 \text{and}\quad \norminf{\{\nabla^2\hmL(\betada^*+\vda)-\nabla^2\hmL(\betada^*)\}\vda}&\le M \Vert{\vda}\Vert_2^2,
 \end{align*}
 where $M=b_3b_1L$.
 \end{enumerate}
\end{prop}

We use the local $\ell_1$-penalized estimator computed on the master machine as the initial estimator. Assume that it satisfies $\widehat{\betada}^{(\mathrm{I},0)}-\betada^*\in\Lambda(\mathtt{g})$ and  $r_0=\normtwo{\widehat{\betada}^{(\mathrm{I},0)}-\betada^*}\lesssim\sqrt{(d+s)\log(p)/n_1}$, which are mild for the Lasso-type estimator \citep{negahban2012unified}. To ensure that the estimators achieve the optimal error rate at each stage, suppose that the iteration numbers $K_{\mathrm{I}}\gtrsim \log(r_0/r^*_{\mathrm{I}})/\log(1/f_{\mathrm{I}})$ and $K_{\mathrm{II}} \gtrsim  \log(r_{\mathrm{I}}^*/r^*_{\mathrm{II}})/\log(1/\overline{f}_{\mathrm{II}})$ as in Condition~\ref{cond:general}. In particular, we observe that the iteration number $K_{\mathrm{I}}$ and $K_{\mathrm{II}}$  scales logarithmically with the global-master ratio $N/n_1$ and doubly logarithmically with the dimensionality $p$, respectively. More iterations are needed to recover the global efficiency as $N/n_1$ grows, since the gap between local and global sample sizes is widened. In the balanced sample case, the global-master ratio equals to the number of data sources $m$. The logarithmic scaling significantly reduces the communication burden and hence allows more data holders to cooperate in a distributed system. In contrast, the DC method generally adopts a one-shot averaging strategy that cannot further reduce the bias inherited from limited local samples, which results in a stringent restriction on the number of data sources $m$.

\begin{cor}\label{thm:oraCST_GLM}
 Suppose Assumptions~\ref{assump:penalty},~\ref{assump:h_end} and~\ref{cond:glm} hold. For any positive sequence $\{t_N\}_N$ diverging to infinity, assume that $\mathbb{E}\normtwo{\boldsymbol{\Omega} \nabla_\bda\ell(\betada^*;\zda_i)}^3 = o(N^{1/2})$, the beta-min condition $|\bgamma^*_{\Sgamma}|\ge(\alpha_0+\alpha_1)\lambda_{\mathrm{II}}$ holds, $(s+d)^3(\log(p)+t_N)=o(n_1)$, and $\normtwo{\hda}^2=o(\sqrt{t_N/N})$. Then we have
 \[
 \sup_{x\ge0}|\mathbb{P}(T^\ora_S\le x)-\mathbb{P}(\chi^2(r,N\hda^\trans \bV^{-1}\hda) \le x)|\to 0.
 \]
\end{cor}

\begin{rmk}
The required assumptions are similar to those in Condition~\ref{cond:general}. The sample scaling $(s+d)^3(\log(p)+t_N)=o(n_1)$ ensures that the contraction factors $f_\mathrm{I}$ and $\overline{f}_{\mathrm{II}}$ are strictly less than $1$, indicating that we can shrinkage the approximation error of the surrogate loss and mitigate biases from high-dimensional components in each iteration. It also implies that the Kiefer-Bahadur representation remainder decays faster than $N^{-1/2}$.  
\end{rmk}

\begin{cor}\label{thm:CST_GLM}
 In addition to the assumptions in Corollary~\ref{thm:oraCST_GLM}, suppose that   $\normtwo{\widehat{\Jda}_0-\Jda_0} =o_p(1)$ and the iteration number in Stage II satisfies $K_{\mathrm{II}}\gtrsim \lceil\log(\widehat{s})\rceil$, where $\widehat{s}=|\mathrm{supp}(\widehat{\bgamma}^\iternum{\mathrm{II},0})|+s$. Then we have
 \[
 \sup_{x\ge0}|\mathbb{P}(T_S\le x)-\mathbb{P}(\chi^2(r,N\hda^\trans \bV^{-1}\hda) \le x)|\to 0.
 \]
\end{cor}

\section{Numerical studies}\label{sec:sim}
In this section, we conduct simulation studies to investigate the finite sample performance of the CST. In particular, we examine the empirical Type I error and power analysis of the CST across different kinds of linear hypotheses and collaborative settings. Moreover, we compare our test with the divide-and-conquer (DC) test proposed by \citet{battey2018distributed} to show that the CST relaxes significantly the strict restriction on the number of
machines with power enhancement, hence allowing more data holders to cooperate in a distributed system. In all simulations, we adopt the SCAD penalty and use the high-dimensional Bayesian information criterion (HBIC) to tune the penalization parameters \citep{wu2020survey}. The stopping criterion for the number of iterations is when the difference between the new and previous estimators is smaller than 0.001, with a maximum of 10 iterations.  The number of Monte Carlo replications is $500$. At the end of this section, we present a real data example.

\subsection{Testing linear hypothesis}\label{sec:sim_linear_hypo}
We consider two models: the linear regression with the standard Gaussian noise and the logistic regression. To generate the covariates, we consider $\xda\sim\mathcal{N}(\zeroda_p,\Sigma)$, where $\Sigma$ is a Toeplitz matrix with $\Sigma_{jk}=0.5^{|j-k|}$. We choose the local sample size $n=200$,  the number of data sources $m\in\{20,50\}$, and the dimensionality $p\in\{1000,10000\}$. {Additional results with $m\in\{10,100\}$ are reported in Appendix~F of the Supplementary Material due to space limitations.} Three types of linear hypotheses are investigated:
 \begin{flalign}
 \text{\textbf{(Univariate hypothesis)}} &&    H^\iternum{1}_0:  \beta_1^* = 0 ; \qquad\qquad\qquad\quad&& \label{eq:uni_hyp}
 \end{flalign}
 \begin{flalign}
\text{\textbf{(Multivariate hypothesis)}} && H^\iternum{2}_0: (\beta_1^*,\beta_2^*,\beta_3^*)^\trans = \zeroda_3 ; \qquad\qquad\qquad\quad&& \label{eq:multi_hyp}
 \end{flalign} 
 \begin{flalign}
\text{\textbf{(Linear hypothesis)}}   &&  H^\iternum{3}_0: \beta_4^*-\beta_5^* = 0 .\quad\quad\quad\quad&& \label{eq:lin_hyp}
 \end{flalign}
We set the true value $\betada^* = (h_1,0,0,1,1,\zeroda_{p-5}^\trans)^\trans$ and $\betada^* = (0,0,0,1+h_2,1,\zeroda_{p-5}^\trans)^\trans$ for the first two hypotheses and the last hypothesis, respectively, where $h_1$ and $h_2$ represent the degree of violation of the null hypothesis.

\begin{figure}[htbp!]
 \centering
 \includegraphics[width=13.2cm]{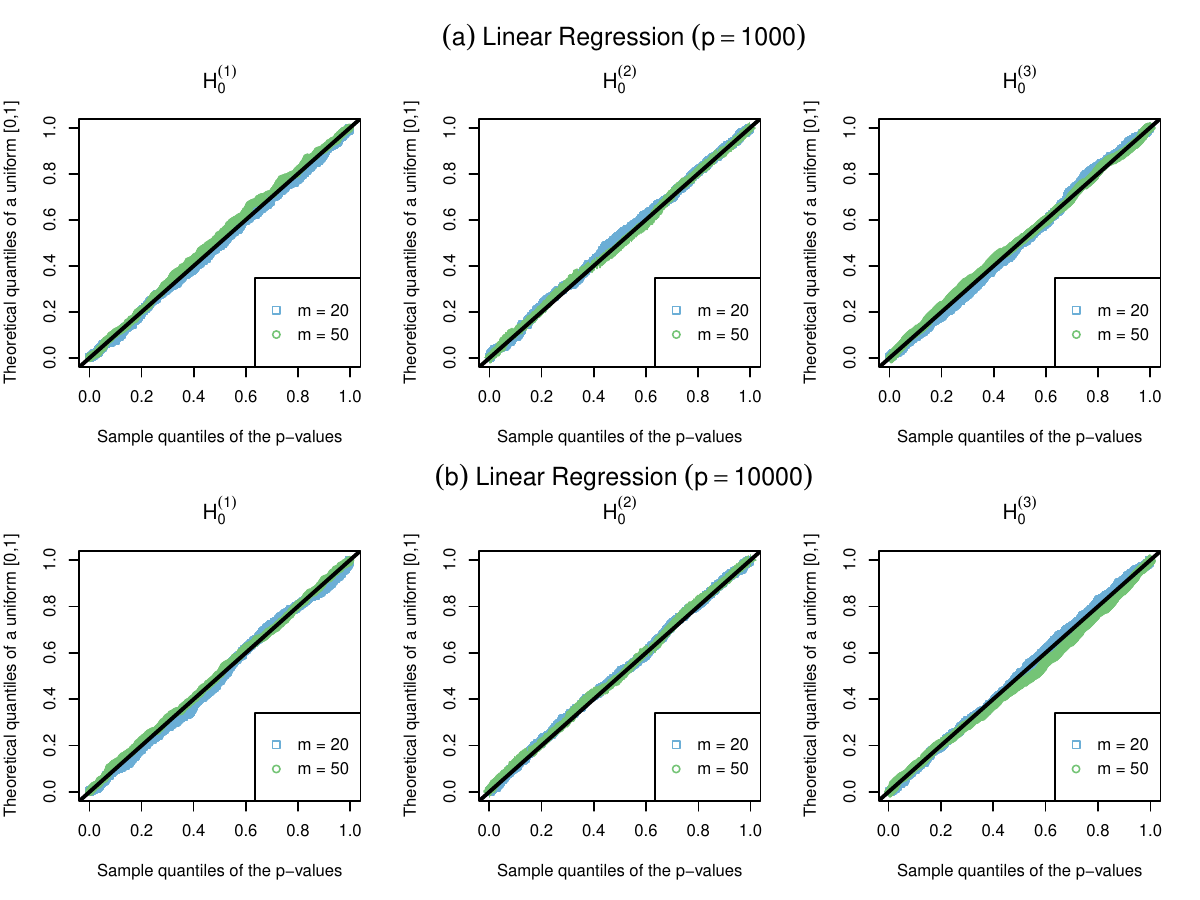}
 \caption{QQ plots of the p-values of the CST  against the theoretical quantiles of the uniform $[0, 1]$ distribution under the null hypothesis in the linear regression setting. The black solid line represents the equal line.}
 \label{fig:linear_QQ}
\end{figure}

{\textbf{Linear regression}.} Simulated data are generated from
$
y=\xda^\trans \betada^* + \epsilon
$, where $\epsilon\sim \mathcal{N}(0,1)$. Table~\ref{tab:linear} reports the empirical rejection probabilities of the CST at different local alternatives $h_1$ and $h_2$ for three types of linear hypotheses. Based on the results, it can be seen that under these null hypotheses, Type I error rates of the CST are well controlled and close to the nominal level $\alpha=0.05$ for all the settings. Under the alternative hypotheses, the powers of the CST increase as $h_1$ or $h_2$ increases, showing the consistency of our testing procedure. Moreover, the CST performs merely same as the OCST, which validates the oracle property. In addition, Figure~\ref{fig:linear_QQ} depicts the QQ plots of the p-values of the CST  against the theoretical quantiles of the uniform $[0, 1]$ distribution under the null hypothesis. The distributions of the p-values are close to uniform, and hence we verify the theoretical result that the CST statistic converges to its limiting distribution.

\begin{figure}[htbp!]
 \centering
 \includegraphics[width=13.2cm]{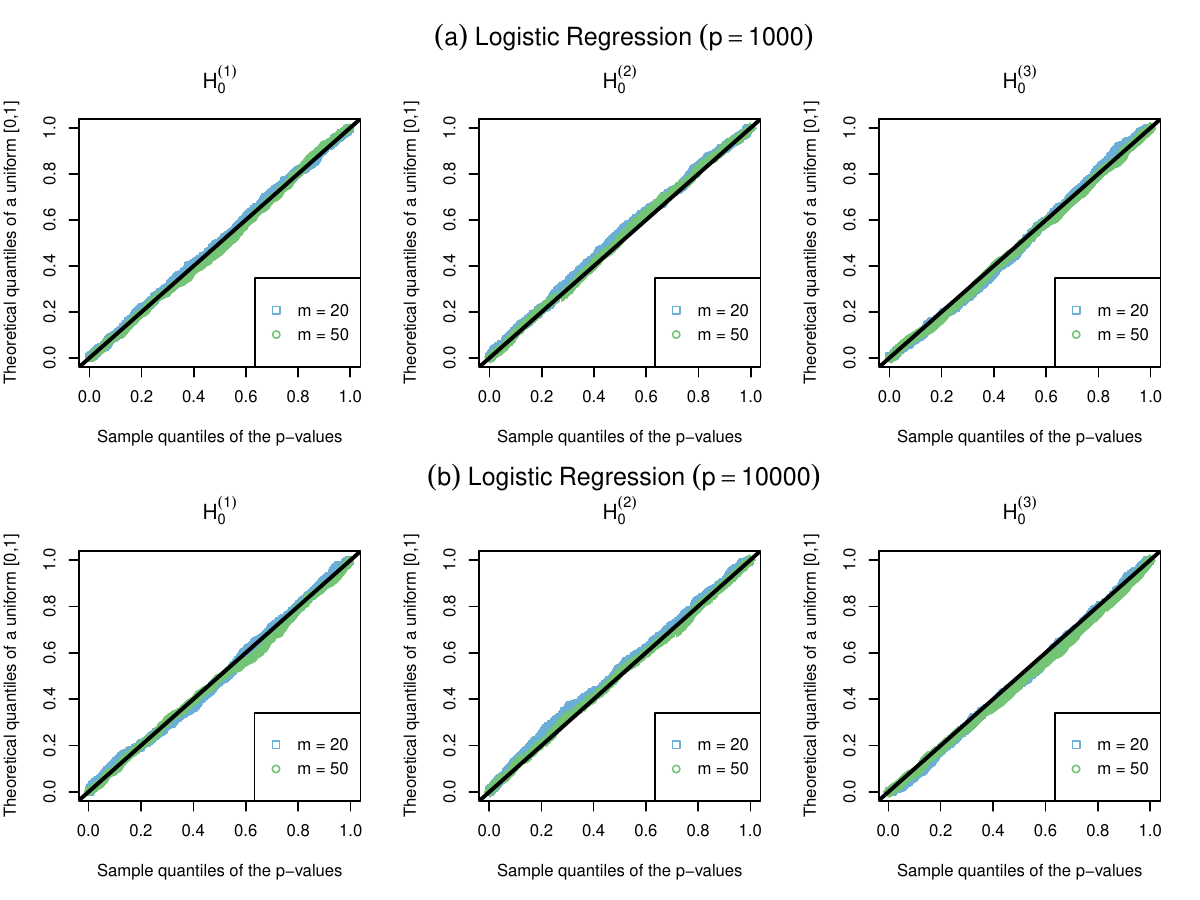}
 \caption{QQ plots of the p-values of the CST against the theoretical quantiles of the uniform $[0, 1]$ distribution under the null hypothesis in the logistic regression setting. The black solid line represents the equal line.}
 \label{fig:logistic_QQ}
\end{figure}

{\textbf{Logistic regression}.} We consider the logistic regression model $
\mathrm{logit}(\mathbb{P}(y=1\mid \xda)) = \xda^\trans\betada^*$, 
where $\mathrm{logit}(p)=\log\{p/(1-p)\}$ is the logit link function. Table~\ref{tab:logistic} reports the empirical rejection probabilities of the CST at different local alternatives $h_1$ and $h_2$ for three types of linear hypotheses. In addition, Figure~\ref{fig:logistic_QQ} depicts the QQ plots of the p-values of the CST against the theoretical quantiles of the uniform $[0, 1]$ distribution under the null hypothesis. The findings are very similar to those in the previous example.

\begin{table}[htbp]
\centering
\caption{Empirical rejection probabilities of the CST and OCST with nominal level $\alpha=0.05$ under the linear regression setting. Results for $n=200$, $m\in\{20, 50\}$, and $p\in\{1000,10000\}$ are reported.  The standard error of estimating rejection rate $0.05$ based on 500 trials is $0.010$.}\label{tab:linear}
     \renewcommand{\arraystretch}{1.2}
 \scalebox{0.95}{
\begin{tabular}{clccclccclccc}
  \hline
  \multicolumn{13}{c}{$p=1000$} \\
\hline
\multirow{7}{*}{$m=20$} & \multicolumn{1}{c}{} & \multicolumn{3}{c}{$H_0^{(1)}$} &  & \multicolumn{3}{c}{$H_0^{(2)}$} &  & \multicolumn{3}{c}{$H_0^{(3)}$} \\
  \cline{3-5}\cline{7-9}\cline{11-13}
  &  & $h_1$ & CST & OCST  &  & $h_1$ & CST & OCST  &  & $h_2$ & CST & OCST  \\
  \cline{3-5}\cline{7-9}\cline{11-13}
  &  & 0 & 0.050 & 0.050 &  & 0 & 0.048 & 0.048 &  & 0 & 0.046 & 0.048 \\
  &  & 0.015 & 0.192 & 0.192 &  & 0.020 & 0.230 & 0.230 &  & 0.030 & 0.210 & 0.210 \\
  &  & 0.030 & 0.598 & 0.598 &  & 0.040 & 0.702 & 0.702 &  & 0.050 & 0.462 & 0.462 \\
  &  & 0.040 & 0.846 & 0.846 &  & 0.050 & 0.884 & 0.884 &  & 0.070 & 0.740 & 0.740 \\
  &  & 0.050 & 0.968 & 0.968 &  & 0.060 & 0.972 & 0.972 &  & 0.090 & 0.93  & 0.930 \\
  \hline
\multirow{7}{*}{$m=50$} & \multicolumn{1}{c}{} & \multicolumn{3}{c}{$H_0^{(1)}$} &  & \multicolumn{3}{c}{$H_0^{(2)}$} &  & \multicolumn{3}{c}{$H_0^{(3)}$} \\
  \cline{3-5}\cline{7-9}\cline{11-13}
  &  & $h_1$ & CST & OCST  &  & $h_1$ & CST & OCST  &  & $h_2$ & CST & OCST  \\
  \cline{3-5}\cline{7-9}\cline{11-13}
  &  & 0 & 0.046 & 0.046 &  & 0 & 0.064 & 0.064 &  & 0 & 0.060 & 0.060 \\
  &  & 0.015 & 0.412 & 0.412 &  & 0.020 & 0.464 & 0.464 &  & 0.030 & 0.414 & 0.420 \\
  &  & 0.030 & 0.926 & 0.926 &  & 0.040 & 0.986 & 0.986 &  & 0.050 & 0.808 & 0.808 \\
  &  & 0.040 & 0.996 & 0.996 &  & 0.050 & 0,998 & 0.998 &  & 0.070 & 0.982 & 0.982 \\
  &  & 0.050 & 1.000 & 1.000 &  & 0.060 & 1.000 & 1.000 &  & 0.090 & 1.000 & 1.000  \\
  \hline
  \multicolumn{13}{c}{$p=10000$} \\
  \hline
\multirow{7}{*}{$m=20$} & \multicolumn{1}{c}{} & \multicolumn{3}{c}{$H_0^{(1)}$} &  & \multicolumn{3}{c}{$H_0^{(2)}$} &  & \multicolumn{3}{c}{$H_0^{(3)}$} \\
  \cline{3-5}\cline{7-9}\cline{11-13}
  &  & $h_1$ & CST & OCST  &  & $h_1$ & CST & OCST  &  & $h_2$ & CST & OCST  \\
  \cline{3-5}\cline{7-9}\cline{11-13}
  &  & 0 & 0.044 & 0.044 &  & 0 & 0.054 & 0.054 &  & 0 & 0.042 & 0.042 \\
  &  & 0.015 & 0.190 & 0.190 &  & 0.020 & 0.204 & 0.204 &  & 0.030 & 0.212 & 0.212 \\
  &  & 0.030 & 0.604 & 0.604 &  & 0.040 & 0.702 & 0.702 &  & 0.050 & 0.484 & 0.484 \\
  &  & 0.040 & 0.850 & 0.850 &  & 0.050 & 0.898 & 0.898 &  & 0.070 & 0.754 & 0.754 \\
  &  & 0.050 & 0.958 & 0.958 &  & 0.060 & 0.978 & 0.978 &  & 0.090 & 0.920 & 0.920 \\
  \hline
\multirow{7}{*}{$m=50$} & \multicolumn{1}{c}{} & \multicolumn{3}{c}{$H_0^{(1)}$} &  & \multicolumn{3}{c}{$H_0^{(2)}$} &  & \multicolumn{3}{c}{$H_0^{(3)}$} \\
  \cline{3-5}\cline{7-9}\cline{11-13}
  &  & $h_1$ & CST & OCST  &  & $h_1$ & CST & OCST  &  & $h_2$ & CST & OCST  \\
  \cline{3-5}\cline{7-9}\cline{11-13}
  &  & 0 & 0.054 & 0.054 &  & 0 & 0.062 & 0.062 &  & 0 & 0.062 & 0.062 \\
  &  & 0.015 & 0.418 & 0.418 &  & 0.020 & 0.481 & 0.481 &  & 0.030 & 0.378 & 0.378 \\
  &  & 0.030 & 0.912 & 0.912 &  & 0.040 & 0.972 & 0.972 &  & 0.050 & 0.832 & 0.832 \\
  &  & 0.040 & 0.994 & 0.994 &  & 0.050 & 0.998 & 0.998 &  & 0.070 & 0.986 & 0.986 \\
  &  & 0.050 & 1.000 & 1.000 &  & 0.060 & 1.000 & 1.000 &  & 0.090 & 1.000 & 1.000  \\
  \hline
\end{tabular}
}
\end{table}

\begin{table}[htbp]
\centering
\caption{Empirical rejection probabilities of the CST and OCST with nominal level $\alpha=0.05$ under the logistic regression setting. Results for $n=200$, $m\in\{20, 50\}$, and $p\in\{1000,10000\}$ are reported. The standard error of estimating rejection rate $0.05$ based on 500 trials is $0.010$.}\label{tab:logistic}
     \renewcommand{\arraystretch}{1.2}
 \scalebox{0.95}{
\begin{tabular}{clccclccclccc}
  \hline
  \multicolumn{13}{c}{$p=1000$} \\
\hline
\multirow{7}{*}{$m=20$} &  & \multicolumn{3}{c}{$H_0^{(1)}$} &  & \multicolumn{3}{c}{$H_0^{(2)}$} &  & \multicolumn{3}{c}{$H_0^{(3)}$} \\
\cline{3-5}\cline{7-9}\cline{11-13}
  &  & $h_1$ & CST & OCST  &  & $h_1$ & CST & OCST  &  & $h_2$ & CST & OCST  \\
  \cline{3-5}\cline{7-9}\cline{11-13}
  &  & 0 & 0.048 & 0.048 &  & 0 & 0.058 & 0.058 &  & 0 & 0.046 & 0.046 \\
  &  & 0.03  & 0.162 & 0.162 &  & 0.03  & 0.106 & 0.106 &  & 0.10  & 0.270 & 0.268 \\
  &  & 0.06  & 0.402 & 0.404 &  & 0.06  & 0.300 & 0.298 &  & 0.15  & 0.560 & 0.560 \\
  &  & 0.09  & 0.738 & 0.738 &  & 0.09  & 0.602 & 0.602 &  & 0.20  & 0.790 & 0.790 \\
  &  & 0.12  & 0.944 & 0.944 &  & 0.12  & 0.860 & 0.860 &  & 0.25  & 0.928 & 0.928 \\
  \hline
\multirow{7}{*}{$m=50$} &  & \multicolumn{3}{c}{$H_0^{(1)}$} &  & \multicolumn{3}{c}{$H_0^{(2)}$} &  & \multicolumn{3}{c}{$H_0^{(3)}$} \\
\cline{3-5}\cline{7-9}\cline{11-13}
&  & $h_1$ & CST & OCST  &  & $h_1$ & CST & OCST  &  & $h_2$ & CST & OCST  \\
\cline{3-5}\cline{7-9}\cline{11-13}
  &  & 0 & 0.044 & 0.044 &  & 0 & 0.042 & 0.042 &  & 0 & 0.054 & 0.054 \\
  &  & 0.03  & 0.252 & 0.252 &  & 0.03  & 0.194 & 0.194 &  & 0.10  & 0.564 & 0.564 \\
  &  & 0.06  & 0.770 & 0.770 &  & 0.06  & 0.622 & 0.622 &  & 0.15  & 0.880 & 0.880 \\
  &  & 0.09  & 0.984 & 0.984 &  & 0.09  & 0.940 & 0.940 &  & 0.20  & 0.992 & 0.992 \\
  &  & 0.12  & 1.000 & 1.000 &  & 0.12  & 0.998 & 0.998 &  & 0.25  & 1.000 & 1.000  \\
  \hline
  \multicolumn{13}{c}{$p=10000$} \\
  \hline
\multirow{7}{*}{$m=20$} &  & \multicolumn{3}{c}{$H_0^{(1)}$} &  & \multicolumn{3}{c}{$H_0^{(2)}$} &  & \multicolumn{3}{c}{$H_0^{(3)}$} \\
  \cline{3-5}\cline{7-9}\cline{11-13}
  &  & $h_1$ & CST & OCST  &  & $h_1$ & CST & OCST  &  & $h_2$ & CST & OCST  \\
 \cline{3-5}\cline{7-9}\cline{11-13}
  &  & 0 & 0.056 & 0.056 &  & 0 & 0.060 & 0.060 &  & 0 & 0.026 & 0.026 \\
  &  & 0.03  & 0.144 & 0.144 &  & 0.03  & 0.122 & 0.122 &  & 0.10  & 0.284 & 0.284 \\
  &  & 0.06  & 0.376 & 0.376 &  & 0.06  & 0.282 & 0.282 &  & 0.15  & 0.568 & 0.568 \\
  &  & 0.09  & 0.730 & 0.730 &  & 0.09  & 0.562 & 0.562 &  & 0.20  & 0.802 & 0.802 \\
  &  & 0.12  & 0.926 & 0.926 &  & 0.12  & 0.856 & 0.856 &  & 0.25  & 0.938 & 0.938 \\
  \hline
\multirow{7}{*}{$m=50$} &  & \multicolumn{3}{c}{$H_0^{(1)}$} &  & \multicolumn{3}{c}{$H_0^{(2)}$} &  & \multicolumn{3}{c}{$H_0^{(3)}$} \\
 \cline{3-5}\cline{7-9}\cline{11-13}
  &  & $h_1$ & CST & OCST  &  & $h_1$ & CST & OCST  &  & $h_2$ & CST & OCST  \\
 \cline{3-5}\cline{7-9}\cline{11-13}
  &  & 0 & 0.042 & 0.042 &  & 0 & 0.060 & 0.060 &  & 0 & 0.040 & 0.036 \\
  &  & 0.03  & 0.272 & 0.272 &  & 0.03  & 0.178 & 0.178 &  & 0.10  & 0.606 & 0.604 \\
  &  & 0.06  & 0.766 & 0.766 &  & 0.06  & 0.614 & 0.614 &  & 0.15  & 0.920 & 0.914 \\
  &  & 0.09  & 0.982 & 0.982 &  & 0.09  & 0.940 & 0.942 &  & 0.20  & 0.990 & 0.990 \\
  &  & 0.12  & 1.000 & 1.000 &  & 0.12  & 0.998 & 0.998 &  & 0.25  & 1.000 & 1.000  \\
  \hline
\end{tabular}
}
\end{table}

\subsection{Comparison with the DC test}\label{subsec:dc}
We compare the proposed CST to the DC test in \citet{battey2018distributed} by allowing more participation of data holders in the distributed system. The DC test uses a statistic $T_{DC} = \sqrt{N} \sum_{k=1}^m \widehat{\beta}_{k,1}^d/(m\widehat{\sigma}_k)$ to test the univariate hypothesis $H_0: \beta_1 = 0$, where $\widehat{\beta}_{k,1}^d$ and $\widehat{\sigma}^2_k$ are the debiased estimator \citep{javanmard2014confidence} and the corresponding variance estimator from each local site, respectively. In the simulations, we fix the dimensionality $p=1000$ and the aggregated sample size $N=3000$, and distribute the full sample into $m$ agents evenly with $n = N/m$. The tests are performed within a linear regression model, with similar results applicable to the logistic regression model as well.

\begin{figure}[htbp]
 \centering
 \adjustbox{center}{\includegraphics[width=16.7cm]{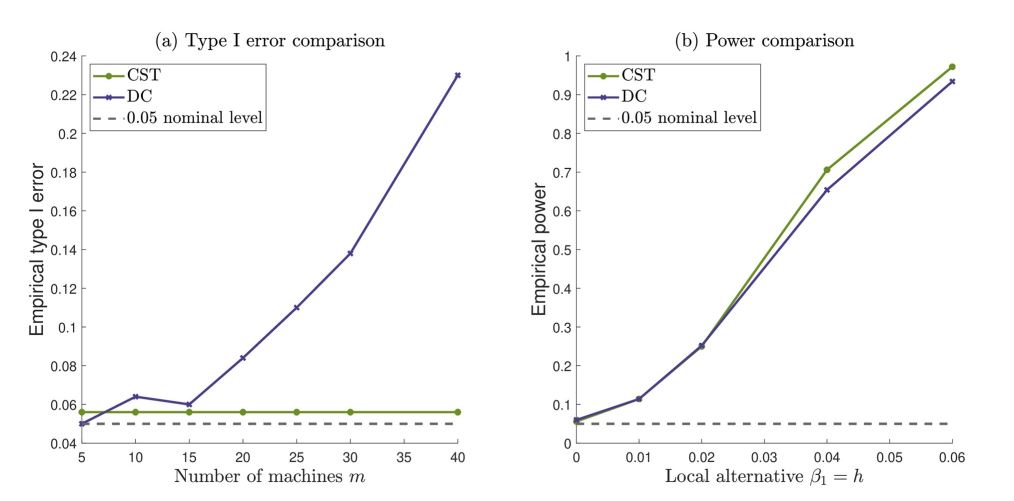}}
 \caption{(a) Empirical type I error of the CST and the DC test \citep{battey2018distributed} when the number of data sites $m$ varies across $\{5,10,15,20,25,30,40\}$. (b) Empirical power of the CST and the DC test under the local alternatives when the number of data sites $m=10$.}
 \label{fig:type1compare}
\end{figure}

\textbf{Effects on the number of data sites $m$.} To examine the validity of each collaborative testing method, we vary the number of data sites $m$ across $\{5,10,15,20,30,40\}$. We set the true parameter $\betada^* = (0,0,0,1,1,\zeroda_{p-5}^\trans)^\trans$ under the null hypothesis. Part (a) of Figure~\ref{fig:type1compare} shows the empirical Type I error of each methods across $m$. It can be seen that when $m\ge20$, the DC test fails to control the Type-I error, whereas the CST performs consistently well across all settings. This is consistent with our theoretical results and demonstrates that our proposed tests are more cooperation-friendly than the DC approach. 

\textbf{Power analysis}. We turn our attention to the power analysis of each collaborative testing method when the number of data sites $m$ is relatively small. We set $m=10$ and the true parameter $\betada^* = (h,0,0,1,1,\zeroda_{p-5}^\trans)^\trans$ under the alternative hypothesis $H_0^a: \beta_1 = h$. Part (b) of Figure~\ref{fig:type1compare} reports the empirical power of each methods across different strength $h$. We find that all the tests are consistent since the power goes to $1$ as the strength $h$ increases. Moreover, the CST exhibits slightly higher power than the DC test, which implies that our approach can be considered more powerful since it circumvents the debiasing procedure. This advantage is supported by a comparison of the asymptotic power function, which is detailed in Appendix~E of the Supplementary Material.

\subsection{Real application}
We apply the CST procedure to a taxi dataset in Chicago, Illinois, which is adapted from the Chicago Open Data Portal\footnote{publicly available at \scriptsize\url{https://data.cityofchicago.org/Transportation/Taxi-Trips-2013-2023-/wrvz-psew}}. The dataset includes quantitative and categorical information about taxi trips in Chicago. Each taxi company is considered a separate data center, with direct data sharing avoided due to possible commercial competitions. Our objective is to analyze the likelihood of tipping using logistic regression collaboratively with non-shared data, focusing on the trip data from the first week of December 2023.

In the data pre-processing, we remove trips with missing information, such as those with zero payment or unknown location, and retain trips from taxis and companies that record more than $1,000$ and $100,000$ trips in 2023, respectively, based on the unique ID numbers. This filtering process leads to $m=10$ companies, representing the top ten service providers in 2023. Then, the dataset for the first week of December 2023 contains $N=67,758$ trips. In the distributed system, the company with the most trips serves as the master site, with $n_1=15,282$ local samples. In addition, quantitative covariates are standardized to have mean zero and variance one, and categorical covariates are transformed into dummy variables, resulting in a total of $p=170$ features (including the intercept). A detailed overview of the covariates is provided in Table~\ref{tab:description_taxi}.

\begin{table}[ht!]
 \centering
\caption{Description for the variables chosen in the Chicago taxi dataset. There are $77$ community areas in Chicago according to the boundary information from the Chicago Open Data Portal. Quantitative covariates are standardized to have mean zero and variance one, and categorical covariates are converted to dummies.}\label{tab:description_taxi}
     \renewcommand{\arraystretch}{1.2}
 \scalebox{0.9}{
 \begin{tabular}{lcp{10cm}}
\hline
\textbf{Response} & & \textbf{Description} \\
\cline{1-1}
Whether a tip is paid && $1$ if a tip is paid, and $0$ otherwise \\
\hline
\textbf{Quantitative covariates} & & \\
\cline{1-1}
Fare & & The fare for the trip (in dollars) \\
Trip Miles & & Distance of the trip (in miles) \\
Trip Seconds & & Time of the trip (in seconds) \\
\hline
\textbf{Categorical covariates} & & \\
\cline{1-1}
Payment Type (PT) & & Type of payment for the trip: Prcard, Cash, Credit Card, and Mobile (with Prcard as the baseline) \\
Day of the Week (DoW) & & The day of the week when the trip started: Monday to Sunday (with Monday as the baseline) \\
Hour of the Day (HoD)& & Four-hour segments according to when the trip started: 2:00-6:00, 6:00-10:00, $\ldots$, and 22:00-2:00 (with 2:00-6:00 as the baseline)\\
Pickup Community Area (\text{PCA})& & The community area where the trip began (with Rogers Park as the baseline) \\
Drop-off Community Area (\text{DCA}) && The community area where the trip ended (with Rogers Park as the baseline) \\
\hline
\end{tabular}}
\end{table}

To formulate the testing hypotheses, we conduct preliminary analysis using taxi data from the last 10 days of November stored at the master site. Specifically, we apply the sure independent ranking and screening \citep[SIRS]{zhu2011model} across all the $170$ covariates using the \texttt{R} package \texttt{VariableScreening} available in \texttt{CRAN}. SIRS ranks the covariates by their correlation with the rank-ordered response. Our analysis focus on the top $8$ and the last $3$ covariates (each with more than one observation for each company), as ranked by SIRS, to test the univariate and multivariate hypotheses, respectively. The results, including the $p$-values of the CST and the DC test, are reported in Table~\ref{tab:pvalue_taxi}. The DC test can be applied only to univariate hypotheses.  For hypotheses where the DC test yields significant $p$-values (i.e., below $0.05$), the CST show even lower $p$-values. Moreover, while the DC test fails to reject the hypothesis `DoW.Thurs = 0', the CST produces a $p$-value of $0.012$. These  results all support that the CST is more powerful than the DC test.  In addition, we find that \{PT.Cash, PT.Credit.Card, \text{PCA}.76\} among the top $8$ covariates show significant impacts on the tipping behavior, with $p$-values below $10^{-10}$. It is interesting to note that \text{PCA}.76 corresponds to the O'Hare community, location of O'Hare International Airport, according to the boundary information\footnote{publicly available at \tiny\url{https://data.cityofchicago.org/Facilities-Geographic-Boundaries/Boundaries-Community-Areas-current-/cauq-8yn6}}. This indicates that trips from the airport are more likely to receive a tip. Moreover, the CST does not reject the multivariate hypothesis $(\text{PCA}.7,\, \text{DCA}.11,\, \text{DCA}.22)^\trans = \zeroda$, suggesting that the last $3$ covariates ranked by SIRS appear to have impacts similar to the baseline community.

\begin{table}[ht!]
 \centering
    \renewcommand{\arraystretch}{1.2}
\caption{The $p$-values based on the CST and the DC test. The top $8$ corvariates \{PT.Credit.Card, PT.Cash, DoW.Thurs, DoW.Tues, \text{PCA}.76, DoW.Wed, DoW.Sat\} and the last $3$ covariates \{\text{PCA}.7, \text{DCA}.11, \text{DCA}.22\} 
are selected by the preliminary ranking and screening procedure. Estimated coefficients for each covariate are also reported based on the unconstrained estimator in Stage II.}\label{tab:pvalue_taxi}
 \scalebox{0.95}{
\begin{tabular}{cccccc}
\hline
Hypothesis & & PT.CreditCard = 0 & PT.Cash = 0 & DoW.Thurs = 0 & DoW.Tues = 0 \\
\cline{3-6}
Coefficient & & $5.600$ & $-4.794$  & $0.154$ & $-0.057$ \\
CST && $<10^{-10}$ & $<10^{-10}$ & $0.012$ & $0.781$ \\
DC && $<10^{-10}$ & $<10^{-10}$ & $0.108$ & $0.958$ \\
\hline
Hypothesis & & DoW.Fri = 0 & \text{PCA}.76 = 0 & DoW.Wed = 0 & DoW.Sat = 0  \\
\cline{3-6}
Coefficient & & $-0.055$ & $0.867$ & $0.123$ & $-0.211$ \\
CST &  & $0.366$ & $<10^{-10}$ & $0.002$ &  $0.001$  \\
DC && $0.499$ & $<10^{-3}$ & $0.042$ & $0.035$ \\
\hline
Hypothesis & & \multicolumn{2}{c}{$(\text{PCA}.7,\, \text{DCA}.11,\, \text{DCA}.22)^\trans = \zeroda$} & \multicolumn{2}{c}{$(\text{PCA}.76,\, \text{DCA}.11,\, \text{DCA}.22)^\trans = \zeroda$} \\
\cline{3-6}
Coefficient & & \multicolumn{2}{c}{$(-0.002,\,0.007,\,-0.070)^\trans$} & \multicolumn{2}{c}{$(0.618,\,-0.012,\,-0.083)^\trans$} \\
CST & &\multicolumn{2}{c}{$0.966$} & \multicolumn{2}{c}{$<10^{-10}$} \\
DC & &\multicolumn{2}{c}{\XSolidBrush} & \multicolumn{2}{c}{\XSolidBrush} \\
\hline
\end{tabular}
}
\end{table}

\section{Discussion}\label{sec:discuss}
There are several directions for future research in this area. We achieve a refined estimation error rate as a by-product through the use of folded-concave penalties, see Appendix~A in the Supplementary Material. These results can be easily applied to the collaborative high-dimensional estimation problems. In addition, our collaborative framework does not cover quantile regression problems due to the non-smooth check loss \citep{he2021smoothed}. Recent work by \citet{tan2022communication} on distributed estimation for smoothed quantile regression suggests a possible extension of the CST procedure to this setting. Moreover, recent federated and multitask learning framework allows certain kind of heterogeneity \citep{liu2024robust,huang2025optimal}, extending the current testing procedure to heterogeneous scenarios would be another intriguing and promising direction.

\bibliographystyle{refstyle}
\bibliography{mybib}

\end{document}